\documentclass[sigconf]{acmart}
\usepackage{subfigure}
\usepackage[norelsize,ruled,vlined,linesnumbered]{algorithm2e}
\usepackage{algorithmic}
\usepackage{url}

\AtBeginDocument{%
  \providecommand\BibTeX{{%
    \normalfont B\kern-0.5em{\scshape i\kern-0.25em b}\kern-0.8em\TeX}}}

\newcommand{\vs}{\vspace{1.5mm}}

\newcommand{\argmax}{\mathop{\rm arg\, max}\limits}
\newcommand{\argmin}{\mathop{\rm arg\, min}\limits}

\setcopyright{none}
\renewcommand\footnotetextcopyrightpermission[1]{}
\begin{document}

\title{Distributed Spatial-Keyword kNN Monitoring for Location-aware Pub/Sub}

\author{Shohei Tsuruoka}
\affiliation{%
\institution{Osaka University}}
\email{tsuruoka.shohei@ist.osaka-u.ac.jp}

\author{Daichi Amagata}
\affiliation{%
\institution{Osaka University}}
\email{amagata.daichi@ist.osaka-u.ac.jp}

\author{Shunya Nishio}
\affiliation{%
\institution{Osaka University}}
\email{nishio.syunya@ist.osaka-u.ac.jp}

\author{Takahiro Hara}
\affiliation{%
\institution{Osaka University}}
\email{hara@ist.osaka-u.ac.jp}

\begin{abstract}
Recent applications employ publish/subscribe (Pub/Sub) systems so that publishers can easily receive attentions of customers and subscribers can monitor useful information generated by publishers.
Due to the prevalence of smart devices and social networking services, a large number of objects that contain both spatial and keyword information have been generated continuously, and the number of subscribers also continues to increase.
This poses a challenge to Pub/Sub systems: they need to continuously extract useful information from massive objects for each subscriber in real time.

In this paper, we address the problem of $k$ nearest neighbor monitoring on a spatial-keyword data stream for a large number of subscriptions.
To scale well to massive objects and subscriptions, we propose a distributed solution, namely D$k$M-SKS.
Given $m$ workers, D$k$M-SKS divides a set of subscriptions into $m$ disjoint subsets based on a cost model so that each worker has almost the same $k$NN-update cost, to maintain load balancing.
D$k$M-SKS allows an arbitrary approach to updating $k$NN of each subscription, so with a suitable in-memory index, D$k$M-SKS can accelerate update efficiency by pruning irrelevant subscriptions for a given new object.
We conduct experiments on real datasets, and the results demonstrate the efficiency and scalability of D$k$M-SKS.
\end{abstract}

\maketitle

\section{Introduction}	\label{section_introduction}
Due to the recent prevalence of GPS-enabled devices, many applications have been generating objects that contain location information and keywords \cite{choudhury2018batch, mahmood2018adaptive}.
They often provide services that retrieve objects useful to users from the generated ones, based on a location-aware publish/subscribe (Pub/Sub) model \cite{hu2015location, li2013location, nishio2020lamps, wang2015ap, wang2016skype}.
In this model, users register queries that specify query locations and keywords as their subscriptions on a Pub/Sub system, and this system delivers appropriate objects generated by publishers (e.g., Point of Interests) to subscriptions based on their query locations and keywords.
It is well known that range and $k$ nearest neighbor ($k$NN) queries support location-aware Pub/Sub systems.
A range query retrieves all objects existing within a user-specified range from a query point, so it cannot control the result size.
This means that users may not obtain any objects or may obtain a huge amount of objects, which is not desirable.
On the other hand, a $k$NN query alleviates this drawback, since users can obtain a reasonable-sized result.
In this paper, hence, we consider $k$NN queries.

\subsection{Motivation}	\label{section_motivation}
In Pub/Sub environments, objects are generated in a streaming fashion, so we have to \textit{continuously update} the $k$NN objects for each subscription.
For example:

\begin{figure}[!t]
	\begin{center}
		\subfigure[At time $t$]{%
			\includegraphics[width=0.485\linewidth]{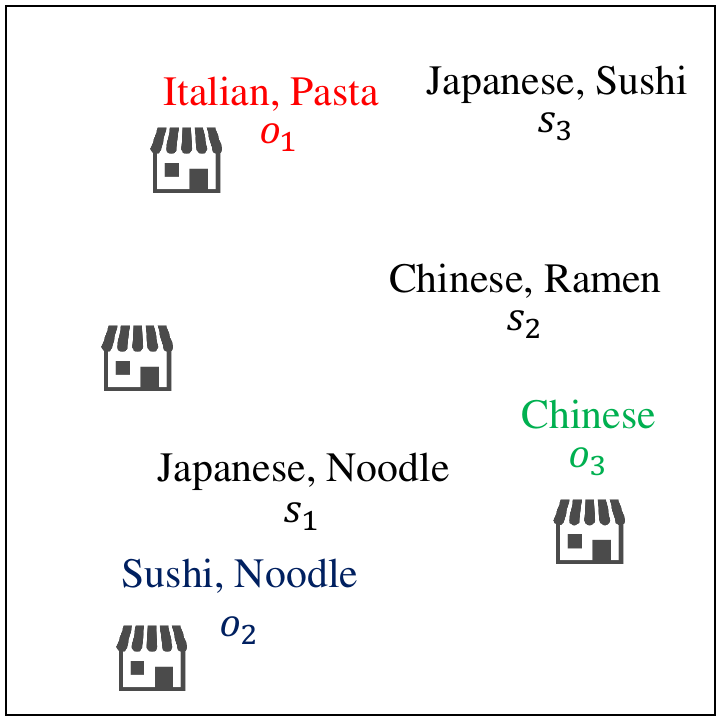}	\label{fig_example-1-a}}
        \subfigure[At time $t + 1$]{%
			\includegraphics[width=0.485\linewidth]{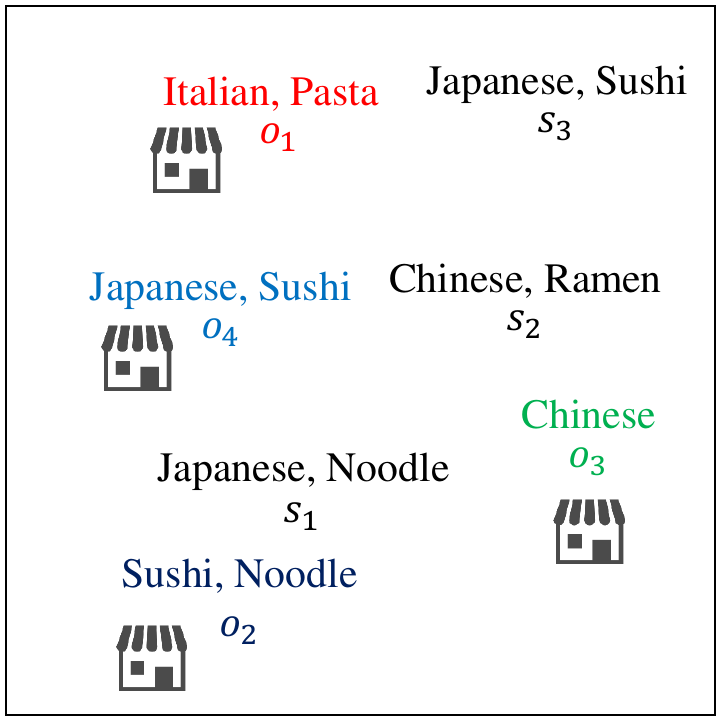}	\label{fig_example-1-b}}
        \caption{An example of $k$NN monitoring in a location-aware Pub/Sub system, where $o_{i}$ and $s_{j}$ respectively denote an spatial-keyword object and a subscription}
        \label{figure_example-1}
	\end{center}
\end{figure}

\vs
\noindent
\textsc{Example 1.}
\textit{Figure \ref{figure_example-1} illustrates an example of $k$NN monitoring in a location-aware Pub/Sub system.
Three subscriptions ($s_{1}$, $s_{2}$, and $s_{3}$) are registered, and the Pub/Sub system monitors $k$NN objects for each subscription.
Assume $k = 1$ and focus on $s_{1}$, which specifies \textsf{Japanese} and \textsf{Noodle} as keywords.
At time $t$, i.e., in Figure \ref{fig_example-1-a}, the NN object for $s_{1}$ is $o_{2}$, because it contains the keyword \textsf{Noodle} and is the nearest to $s_{1}$ among $\{o_{1}, o_{2}, o_{3}\}$.
Also, the NN object for $s_{2}$ ($s_{3}$) is $o_{3}$ ($o_{2}$).
Assume further that a new object $o_{4}$ is generated at time $t + 1$, as shown in Figure \ref{fig_example-1-b}.
Since $o_{4}$ also contains the keyword \textsf{Japanese}, the NN object of $s_{3}$ is updated to $o_{4}$ (and the NN objects for the other subscriptions do not change).}

\vs
\noindent
Users require up-to-date results, so Pub/Sub systems have to efficiently update $k$NN objects of their subscriptions when new objects are given.
However, this is a difficult task, because many applications employing Pub/Sub systems have to deal with a lot of (often million-scale) subscriptions \cite{wang2015ap_}.
Besides, due to the usefulness of location-aware Pub/Sub systems, the number of subscriptions is further increasing \cite{wang2017top}.
It is therefore hard for a single server to update the result for each subscription in real time \cite{chen2017distributed}.
This suggests that we need to make location-aware Pub/Sub systems efficient and scalable, motivating us to consider a distributed solution:
given multiple workers, each registered subscription is assigned to a specific worker so that parallel $k$NN update is enabled.

\vs
\noindent
\textbf{Challenge.}
Although a distributed solution is promising, it has some challenges to scale well to massive objects and subscriptions (i.e., continuous spatial-keyword $k$NN queries).

(1) A distributed solution has to maintain load balancing.
This is not trivial for continuous spatial-keyword $k$NN queries, because each subscription specifies arbitrary locations and keywords., i.e., the loads of subscriptions are different and not explicitly provided.

(2) It is necessary to deal with subscription insertions and deletions.
Although some variants of the spatial-keyword $k$NN monitoring problem \cite{hu2015location, wang2016skype} accept subscription insertions and deletions, these solutions consider centralized environments and extending them for decentralized environments is not trivial.
In addition, \cite{chen2017distributed, wang2017top} assume subscription insertions and deletions in distributed processing environments.
However, \cite{chen2017distributed} considers not the costs of subscriptions but the number of them, which is not effective for load balancing, and \cite{wang2017top} does not consider load balancing.

\subsection{Contribution}
We overcome these challenges and propose two baselines and D$k$M-SKS (\underline{D}istributed \underline{$k$}NN \underline{M}onitoring on \underline{S}patial-\underline{K}eyword data \underline{S}tream).
Our solutions employ
\begin{itemize}
	\setlength{\leftskip}{-4.0mm}
    \item	\textbf{Cost models for subscriptions}:
    		We design cost models for subscriptions, so that we can estimate the load of a given subscription when a new object is generated.
    		Specifically, we propose keyword- and space-oriented cost models.
            Our models use a practical assumption and can deal with new subscriptions.
            Based on these models, we further propose a hybrid of these two models.
    \item	\textbf{Cost-based subscription partitioning}:
    		Based on our cost models, a set of subscriptions is divided into disjoint subsets, each of which is assigned to a specific worker.
            In particular, D$k$M-SKS considers both spatial and keyword information, so that $k$NN update costs can be minimized.
            We use a greedy algorithm for subscription partitioning, because optimal cost-based partitioning is NP-hard.
\end{itemize}
Furthermore, D$k$M-SKS allows an arbitrary exact algorithm for $k$NN update.
This is a good property because it can implement a state-of-the-art to accelerate performance.
To demonstrate the efficiency of D$k$M-SKS, we conduct experiments on two real datasets.
From the experimental results, we confirm that D$k$M-SKS outperforms the baselines and a state-of-the-art technique.
This is the full version of our preliminary paper \cite{tsuruoka2020distributed}.

\vs
\noindent
\textbf{Organization.}
The rest of this paper is organized as follows.
We formally define our problem in Section \ref{section_preliminary}.
Then, we design baseline solutions in Section \ref{section_baseline}.
We propose D$k$M-SKS in Section \ref{section_propose}, and introduce our experimental results in Section \ref{section_experiment}.
Related works are reviewed in Section \ref{section_related-work}.
Finally, this paper is concluded in Section \ref{section_conclusion}.

\section{Preliminary}	\label{section_preliminary}
\underline{\textbf{Problem definition.}}
Let us first define spatial-keyword objects.

\vs
\noindent
\textsc{Definition 1 (Spatial-keyword object).}
\textit{A spatial-keyword object $o$ is defined as $o = \langle p,\psi,t\rangle$, where $p$ is a 2-dimensional location of $o$, $\psi$ is a set of keywords held by $o$, and $t$ is the time-stamp when $o$ is generated.}

\vs
\noindent
Without loss of generality, hereinafter, we call $o$ \textit{object} simply.
Note that we assume discrete time in this paper.
Next, we define continuous spatial-keyword $k$ nearest neighbor ($k$NN) queries.

\vs
\noindent
\textsc{Definition 2 (Continuous spatial-keyword $k$NN query).}
\textit{A continuous spatial-keyword $k$NN query $s$ is defined as $s = \langle p,\psi,k,t\rangle$, where $p$ is a 2-dimensional location of interest for $s$, $\psi$ is a set of keywords in which $s$ is interested, $k$ is the number of results required by $s$, and $t$ is the time-stamp when $s$ is registered.
Let $O$ be a set of objects generated so far, and let $O(s)$ be the set of objects $o \in O$ where $o.\psi \cap s.\psi \neq \varnothing$ and $s.t \leq o.t$.
Given $O(s)$, this query monitors a set of objects $A$ that satisfy (i) $|A| = k$ and (ii) $\forall o \in A$, $\forall o' \in O(s) - A$, $dist(o.p,s.p) \leq dist(o'.p,s.p)$, where $dist(p,p')$ evaluates the Euclidean distance between points $p$ and $p'$ (ties are broken arbitrarily).}

\vs
\noindent
That is, we consider continuous spatial-keyword $k$NN queries with a \textit{boolean} (i.e., \textsf{OR}) semantic for keywords \cite{almaslukh2018evaluating, amagata2015distributed, chen2013spatial} and a time constraint \cite{amagata2016diversified, qiao2016range} for obtaining fresh objects as much as possible.
A subscription is corresponding to a continuous spatial-keyword $k$NN query in this paper, as shown in Example 1.
We hence use them interchangeably.
Then, our problem is defined as follows:

\begin{figure}[!t]
	\begin{center}
    	\includegraphics[width=0.99\linewidth]{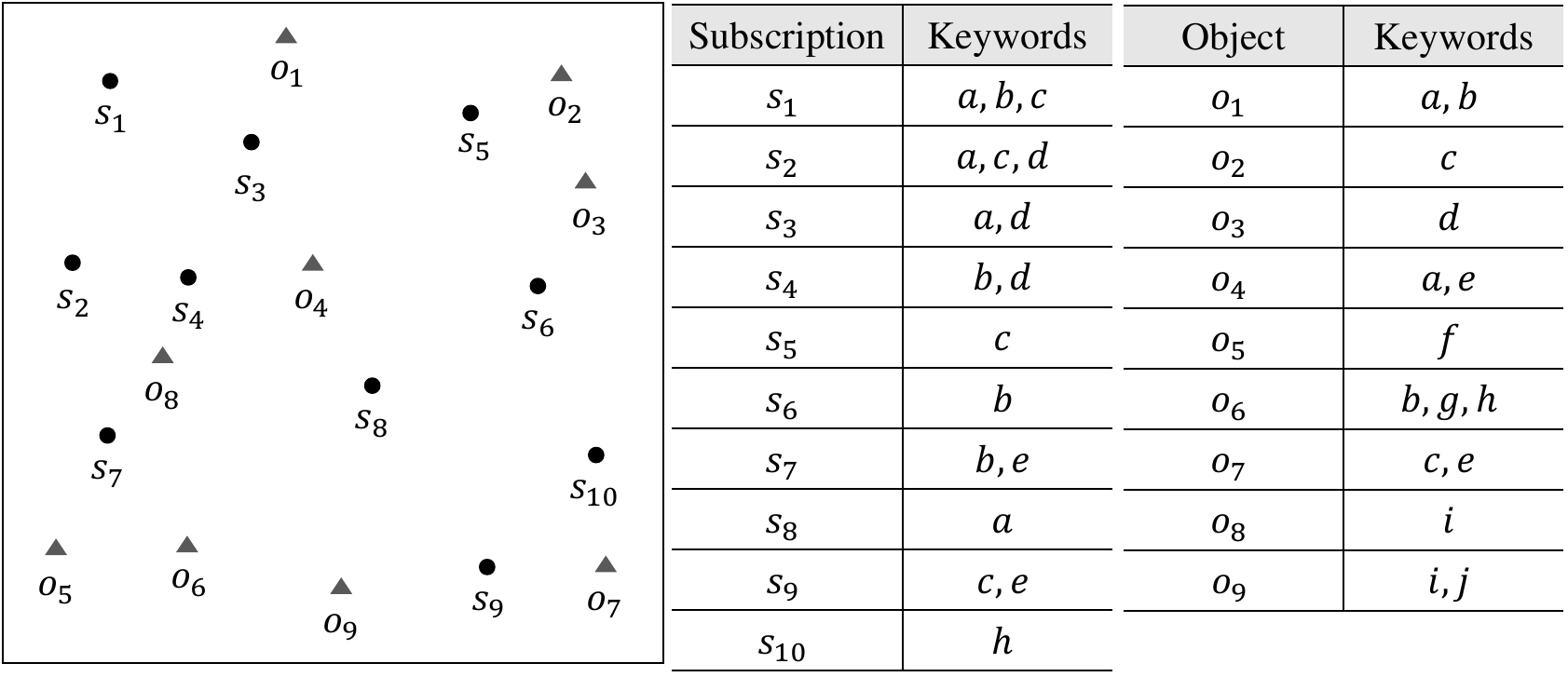}
        \caption{A toy example of objects and subscriptions that have been respectively generated and registered at time $t$}
        \label{figure_example-2}
	\end{center}
\end{figure}

\vs
\noindent
\textsc{Problem statement.}
\textit{Given $O$ and a set of registered subscriptions $S$, our problem is to exactly monitor $A$ for each subscription $\in S$.}

\vs
\noindent
\textsc{Example 2.}
\textit{Figure \ref{figure_example-2} illustrates a toy example which is used throughout this paper.
Assume that $s_{1}$, ..., $s_{10}$ ($o_{1}, ..., o_{9}$) have been registered (generated) at time $t$.
Consider $s_{1}$, then $O(s_{1}) = \{o_{1},o_{4},o_{6},o_{7}\}$, because they contain \textsf{a}, \textsf{b}, or \textsf{c}.
Assuming $s_{1}.k = 2$, $A$ of $s_{1}$ is $\{o_{1},o_{4}\}$.}

\vs
\noindent
This paper proposes a distributed solution to achieve real-time monitoring and scale well to large $|O|$ and $|S|$.

\vs
\noindent
\underline{\textbf{System overview.}}
We assume that a location-aware Pub/Sub system employs a general distributed setting consisting of a main server and $m$ workers \cite{amagata2018space, luo2014distributed}.
The main server (each worker) directly communicates with workers (the main server).
(A worker can be a CPU core or a machine that can use a thread.)
The main server takes the following roles: it
\begin{itemize}
	\setlength{\leftskip}{-4.0mm}
    \item	assigns each subscription to a specific worker,
    \item	receives a stream of objects and broadcasts them to all workers, and
    \item	accepts subscription insertions and deletions.
\end{itemize}
The main operations of each worker are as follows: it
\begin{itemize}
	\setlength{\leftskip}{-4.0mm}
    \item	accepts subscriptions assigned by the main server,
    \item	removes requested subscriptions, and
    \item	updates the $k$NN objects for each assigned subscription.
\end{itemize}

We see that $k$NN objects for each subscription are updated in parallel, thereby this approach is promising for massive subscriptions.
An important problem to achieve this is load balancing.
That is, distributed solutions have to consider how to make the computation time of each worker almost equal when new objects are generated.
We below analyze this problem theoretically.

Let $C(s)$ be the $k$NN update cost of a subscription $s$ (how to obtain $C(s)$ is introduced later).
Furthermore, let $C(w_{i})$ be the cost (load) of a worker $w_{i}$, which is defined as
\begin{equation*}
	C(w_{i}) = \sum_{s \in S(w_{i})}C(s),
\end{equation*}
where $S(w_{i})$ is a set of subscriptions assigned to $w_{i}$.
We want to optimize the load difference between workers with a good subscription assignment.
This can be formalized as follows:

\vs
\noindent
\textsc{Definition 3 (Optimal subscription assignment problem).}
\textit{Given a set of objects $O$, a set of subscriptions $S$, and $m$ workers, this problem is to find a subscription assignment that minimizes}
\begin{equation*}
	\max_{i \in [1,m]}C(w_{i}) - \min_{j \in [1, m]}C(w_{j}).
\end{equation*}

\vs
\noindent
We have the following theorem w.r.t. the above problem \cite{amagata2019identifying}.

\vs
\noindent
\textsc{Theorem 1.}
\textit{The optimal subscription assignment problem is NP-hard.}

\vs
\noindent
It can be seen, from this theorem, that it is not practical to obtain the optimal assignment, which suggests that we need a heuristic approach.
We hence consider $C(s)$ to capture the load of $s$ and then design an approach that partitions $S$ into $m$ disjoint subsets whose loads are well balanced.
Note that $C(s)$ is dependent on a given cost model.
In addition, we consider how to manage new subscriptions (we can easily deal with subscription deletions: the main server simply requests them to the corresponding workers).

\section{Baselines}	\label{section_baseline}
Because this is the first work that proposes a distributed solution for processing continuous spatial-keyword $k$NN queries defined in Definition 2, we first design baseline solutions.
We propose two baselines that respectively employ keyword- and space-oriented subscription partitioning.
We assume that some subscriptions are registered at the initial time, and we partition $S$ when $O$ becomes sufficiently large.
(This is common to D$k$M-SKS.)
We use $O_{init}$ to denote the set of objects when $S$ is partitioned.

\subsection{Keyword-oriented Partition}	\label{section_keyword}
To start with, we design keyword-oriented partition.
One possible approach partitions $S$ so that a set of distinct keywords of the subscriptions held by each worker can be disjoint between workers.
This approach is not efficient, because it does not consider keyword frequencies.
In other words, if a worker has subscriptions with keywords that are contained by many objects, its load becomes heavy, rendering load imbalance.
Hence our keyword-oriented partition takes keyword frequencies into account.

\vs
\noindent
\underline{\textbf{Cost estimation.}}
Similar to \cite{wang2015ap}, we estimate the load of a subscription based on the distributions of keywords in $O_{init}$, because the distributions of large datasets rarely change in practice \cite{yoon2019nets}.
Assume that the appearance probability of each keyword is independent.
Given an object $o$, the probability that a keyword $\lambda$ is contained in $o.\psi$, $P(\lambda)$, is
\begin{equation*}
	P(\lambda) = \frac{|O_{\lambda}|}{|O_{init}|},
\end{equation*}
where $O_{\lambda} \in O_{init}$ is a set of objects $o_{i}$ such that $\lambda \in o_{i}.\psi$.
Recall that $O(s)$ is a set of objects $o_{j}$ where $o_{j}.\psi \cap s.\psi \neq \varnothing$.
Therefore, the $k$NN update cost of a subscription $s$, $C(s)$, can be estimated as:
\begin{equation}
	C(s) = \sum_{\lambda \in s.\psi}P(\lambda).	\label{equation_keyword-cost}
\end{equation}

\vs
\noindent
\underline{\textbf{Subscription partition.}}
Keyword-oriented partition employs the cost model defined in Equation (\ref{equation_keyword-cost}) and a 3/2-approximation greedy algorithm \cite{graham1969bounds} for subscription partitioning.
This approach first computes $C(s)$ for every $s \in S$, and sorts $S$ in descending order of $C(s)$.
Then, this approach sequentially accesses subscriptions while assigning an accessed subscription to $w_{i}$ with the minimum $C(w_{i})$.
Algorithm \ref{algo_assignment} details this approach.

\vs
\noindent
\underline{\textbf{$k$NN update.}}
Each worker $w$ maintains an inverted file $w.I$ to index its assigned subscriptions.
The inverted file is a set of postings lists $w.I[\lambda]$ that maintain subscriptions containing a keyword $\lambda$.
Given a new object $o$ (broadcast by the main server), each worker $w$ computes subscriptions that contain keywords in $o.\psi$, from $w.I$, while pruning irrelevant subscriptions.
After that, $w$ updates the $k$NN of the corresponding subscriptions.

\begin{figure}[!t]
	\begin{center}
    	\includegraphics[width=0.99\linewidth]{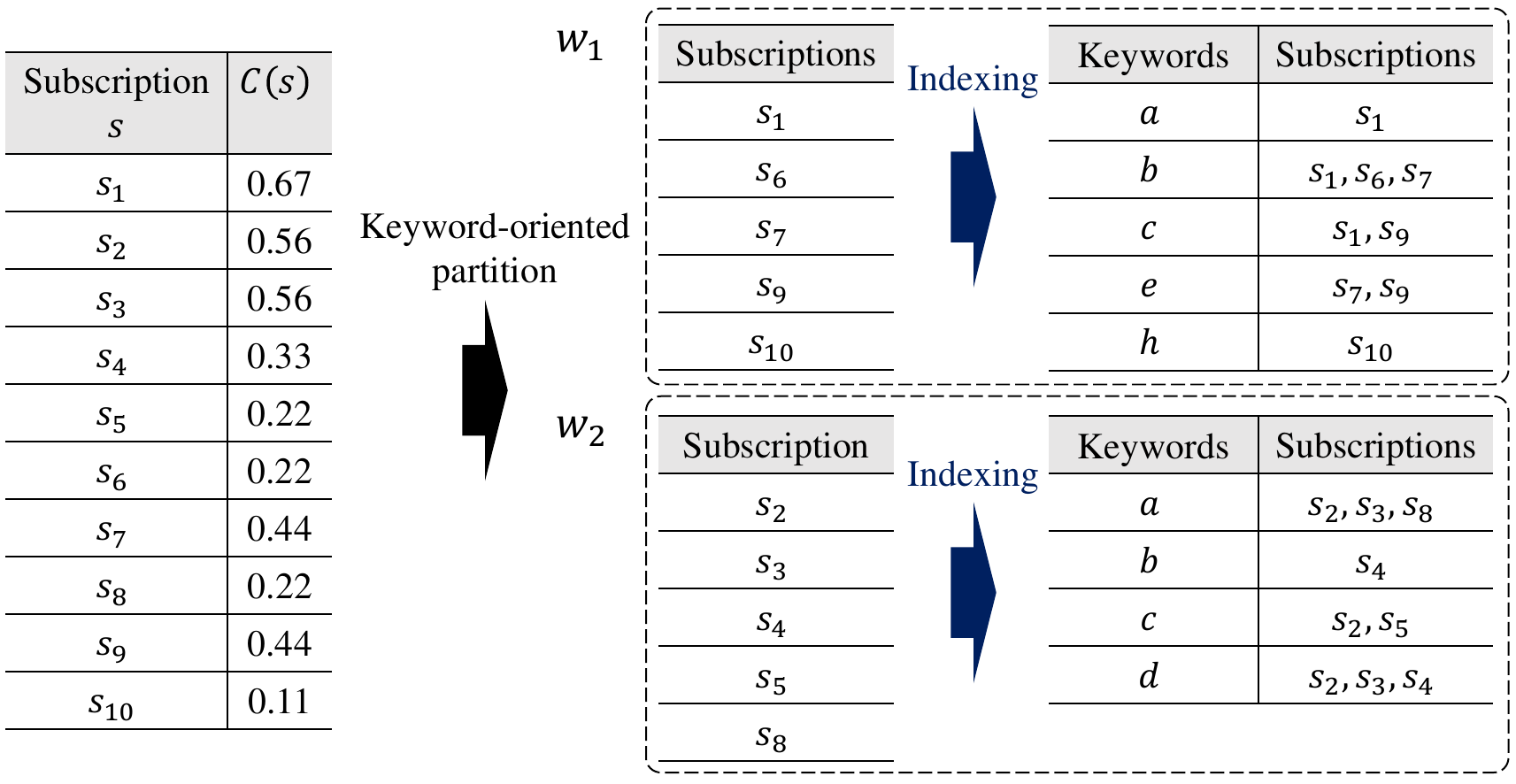}
        \caption{An example of keyword-oriented partition for two workers $w_{1}$ and $w_{2}$ (based on objects and subscriptions in Figure \ref{figure_example-2})}
        \label{figure_example-3}
	\end{center}
\end{figure}

\vs
\noindent
\textsc{Example 3.}
\textit{We partition $S$ in Figure \ref{figure_example-2} into two disjoint subsets for workers $w_{1}$ and $w_{2}$, based on keyword-oriented partition.
Figure \ref{figure_example-3} illustrates an overview.
The left part shows subscriptions and their costs obtained from Equation (\ref{equation_keyword-cost}), and the right part shows the partition result, i.e., $w_{1}$ has $s_{1}$, $s_{6}$, $s_{7}$, $s_{9}$, and $s_{10}$ while $w_{2}$ has $s_{2}$, $s_{3}$, $s_{4}$, $s_{5}$, and $s_{8}$.
They are maintained by inverted files (the most right tables).}

\begin{algorithm}[!t]
	\caption{\textsc{Subscription-Assignment}}
    \label{algo_assignment}
	\DontPrintSemicolon
	\KwIn {$S$ (a set of subscriptions) and $m$ workers}
	Sort $S$ in descending order of cost\;
    Set $C(w) = 0$ for each worker\;
    \For {each $s \in S$}{
    	$w \leftarrow \argmin_{m}C(w)$\;
        $S(w) \leftarrow S(w) \cup \{s\}$, $C(w) \leftarrow C(w) + C(s)$
    }
\end{algorithm}

\vs
\noindent
\underline{\textbf{Subscription insertion.}}
A new subscription $s'$ also can obtain its estimated cost from Equation (\ref{equation_keyword-cost}), because its cost model assumes that the keyword distribution rarely changes \cite{wang2015ap_}.
The main server maintains $C(w)$ for each worker $w$.
(This is common to all of our solutions.)
Given a new subscription $s'$, the main server computes $C(s)$ from Equation (\ref{equation_keyword-cost}).
Then the main server assigns $s'$ to the worker with the minimum $C(w)$.

\vs
\noindent
\underline{\textbf{Subscription deletion.}}
For subscription deletion, the main server simply requests the worker, which has the corresponding subscription, to remove it, then updates $C(w)$.
This is also common to our solutions.

\subsection{Space-oriented Partition}	\label{section_space}
We next design space-oriented partition.
The most straightforward approach is to partition the data space into $m$ equal-sized subspaces.
Clearly, this is not efficient, because some of them have more objects than the others, which also provides load imbalance.
We hence consider a space-based cost model below.

\vs
\noindent
\underline{\textbf{Cost estimation.}}
Consider a set $S_{r}$ of subscriptions that exist in a subspace $r$, and let $C(S_{r})$ be its cost.
Note that $C(S_{r})$ can be a probability that $k$NN objects of subscriptions in $S_{r}$ may be updated, given a new object $o$.
Let $o_{k}$ be the current $k$-th nearest neighbor object of a subscription $s$.
Furthermore, let $B(s)$ be a ball whose center and radius are respectively $s.p$ and $dist(o_{k}.p,s.p)$.
We see that new objects that are generated within $B(s)$ may become new $k$NN of $s$.
Now consider a rectangle $R$ that encloses all balls of $S_{r}$.
It is also true that new objects that are generated within $R$ may become new $k$NNs of $s \in S_{r}$.

The space-based cost also utilizes the distribution of $O_{init}$.
Given a set $O_{R}$ of objects existing within $R$, the probability that a new object is generated within $R$, $P(R)$, is
\begin{equation}
	P(R) = \frac{|O_{R}|}{|O_{init}|}.	\label{equation_space-prob}
\end{equation}
Then we define $C(S_{r})$ as follows:
\begin{equation}
	C(S_{r}) = P(R) \cdot |S_{r}|	\label{equation_space-cost}
\end{equation}
It can be seen that $C(S_{r})$ takes the number of subscriptions into account.
Assume that $R$ is small but contains many subscriptions.
We see that the $k$NN update cost of $R$ is not small when a new object is generated within $R$.
However, without $|S_{r}|$, $C(S_{r})$ is small, which contradicts the above intuition.
We therefore make Equation (\ref{equation_space-cost}) an expected value, different from Equation (\ref{equation_keyword-cost}).

\vs
\noindent
\underline{\textbf{Subscription partition.}}
Here, we introduce how to obtain $R$ (or $r$).
Let $\mathbb{R}^{2}$ be the space where objects and subscriptions exist.
We partition $\mathbb{R}^{2}$ in a similar way to quadtree \cite{finkel1974quad}, motivated by a recent empirical evaluation on a spatial-keyword stream that confirms the superiority of quadtree-based space partition \cite{almaslukh2018evaluating}.
Specifically, we partition $\mathbb{R}^{2}$ into four equal-sized subspaces and compute $C(S_{r})$ for each subspace $r$.
Then we pick the subspace that has the largest $C(S_{r})$ and partition it in the same way.
This is repeated until we have $n \geq \theta \cdot m$, where $n$ and $\theta$ are the number of subspaces and a threshold (system parameter), respectively.

Now we have $n$ disjoint subsets of $S$ and determine their assignment in a similar way to Algorithm \ref{algo_assignment}.
Note that space-oriented partition considers the assignment of subsets $S_{r}$, different from keyword-oriented partition.
That is, the input of the greedy algorithm is a collection of subsets $S_{r}$.

\vs
\noindent
\underline{\textbf{$k$NN update.}}
Space-oriented partition takes a different approach from keyword-oriented partition.
Assume that a worker $w$ has a collection $S(w)$ of $S_{r}$.
For each $S_{r} \in S(w)$, we build an inverted file $I(S_{r})$ for $S_{r}$.
This aims at pruning irrelevant subscriptions, i.e., we can prune $S_{r}$ when a new object is generated within $R$ but does not contain any keywords in $S_{r}$.

Given a new object $o$ that is generated within $R$, $w$ computes subscriptions that contain the keywords in $o.\psi$ by using $I(S_{r})$.
If there are such subscriptions, $w$ updates their $k$NNs.

\begin{figure}[!t]
	\begin{center}
		\subfigure[Space-oriented partition for $S$ in Figure \ref{figure_example-2}]{%
			\includegraphics[width=0.485\linewidth]{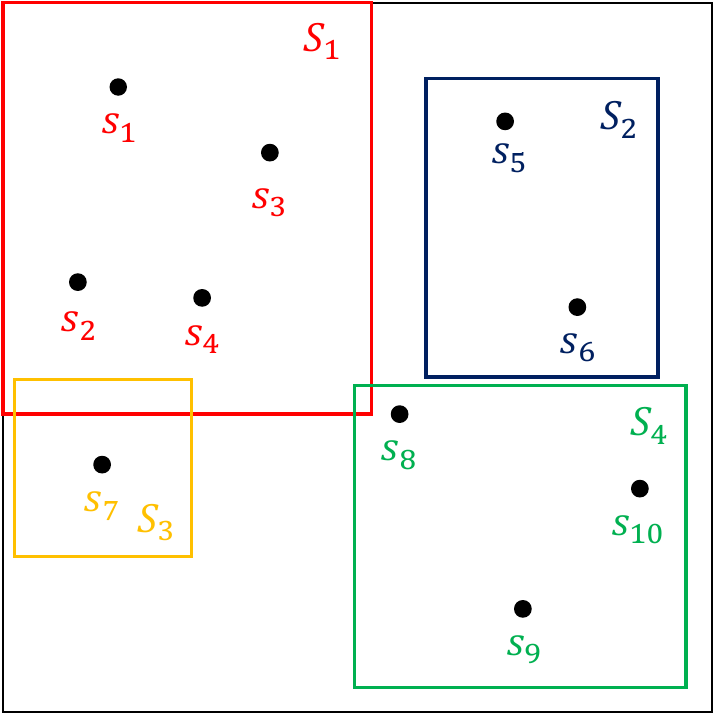}	\label{fig_example-4-a}}
        
        \subfigure[Subscriptions that are assigned to $w_{1}$]{%
			\includegraphics[width=0.485\linewidth]{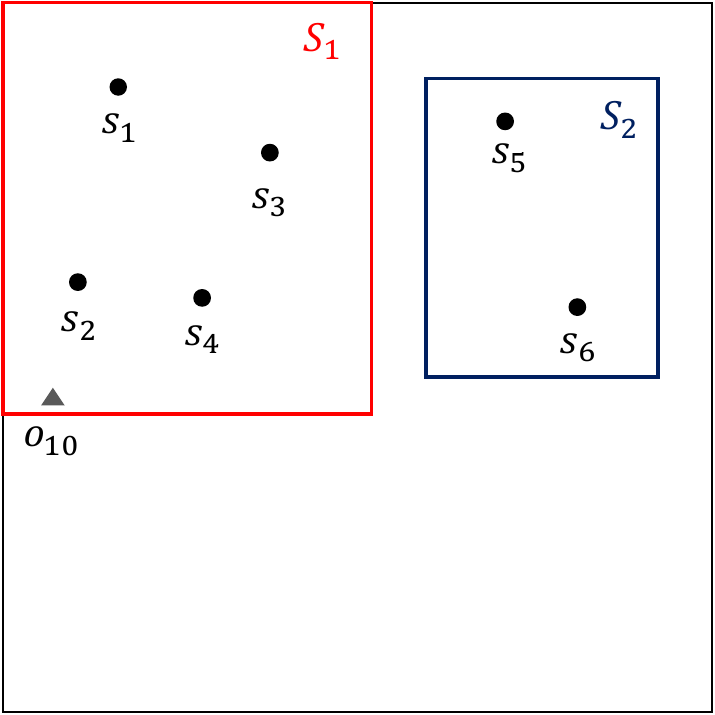}	\label{fig_example-4-b}}
        \subfigure[Subscriptions that are assigned to $w_{2}$]{%
			\includegraphics[width=0.485\linewidth]{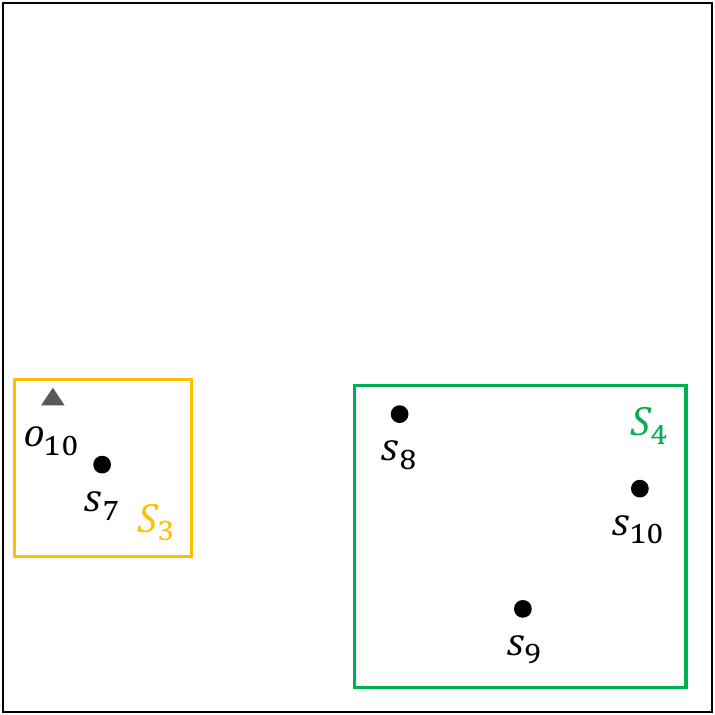}	\label{fig_example-4-c}}
        \caption{An example of space-oriented partition for two workers $w_{1}$ and $w_{2}$ (based on objects and subscriptions in Figure \ref{figure_example-2})}
        \label{figure_example-4}
	\end{center}
\end{figure}

\vs
\noindent
\textsc{Example 4.}
\textit{We partition $S$ in Figure \ref{figure_example-2} into two disjoint subsets for workers $w_{1}$ and $w_{2}$, based on space-oriented partition.
Figure \ref{figure_example-4} illustrates an example.
For simplicity, $S$ is partitioned into four subsets $S_{1}$, $S_{2}$, $S_{3}$, and $S_{4}$, see Figure \ref{fig_example-4-a}.
Assume that their costs are 0.36, 0.11, 0.2, and 0.28, respectively.
Then $S_{1}$ and $S_{2}$ ($S_{3}$ and $S_{4}$) are assigned to $w_{1}$ ($w_{2}$), as shown in Figure \ref{fig_example-4-b} (\ref{fig_example-4-c}).}

\textit{Given a new object $o_{10}$ that is shown in Figures \ref{fig_example-4-b} and \ref{fig_example-4-c}, $w_{1}$ needs to deal with $s_{1}$, $s_{2}$, $s_{3}$, and $s_{4}$, because $o_{11}$ exists within the rectangle of $S_{1}$.
Similarly, $w_{2}$ needs to consider $s_{7}$.}

\vs
\noindent
\underline{\textbf{New subscription.}}
Our space-oriented partition provides a cost with a \textit{set} of subscriptions.
On the other hand, for new subscriptions, we should provide their respective costs, because the number of new subscriptions at a given time is much smaller than that of the initial set of subscriptions.
We therefore estimate the cost of a new subscription based on Equation (\ref{equation_space-prob}).

Given a new subscription $s$, the main server computes its $k$NN among $O_{init}$ to obtain $B(s)$.
(Note that its \textit{exact} $k$NN is monitored after $s$ is assigned to a worker, as $O_{init}$ is not qualified for $O(s)$.)
Then the main server has a rectangle $R$ (i.e., a space $r$) that encloses $B(s)$.
Now we can obtain its cost from Equation (\ref{equation_space-prob}), because $S_{r} = \{s\}$, i.e., Equation (\ref{equation_space-cost}) becomes Equation (\ref{equation_space-prob}).
How to assign $s$ to a worker is the same as keyword-oriented partition.

Although the above approach can deal with new subscriptions, it loses the property of ``space-oriented'', because a new subscription $s$ may be assigned to a worker $w$ that does not have subscriptions close to $s$.
This case may degrade the pruning performance, because the data space, where $w$ has to care, becomes larger.
For example, assume that a new subscription $s_{11}$ is registered and its location is a point in $S_{2}$ of Figure \ref{fig_example-4-a}.
Assume furthermore that $s_{11}$ is assigned to $w_{2}$, then the space, where $w_{2}$ has to take care, becomes larger.

\section{D$k$M-SKS}	\label{section_propose}
\begin{figure}[!t]
	\begin{center}
        \subfigure[A distribution in which space-oriented partition is effective]{%
			\includegraphics[width=0.485\linewidth]{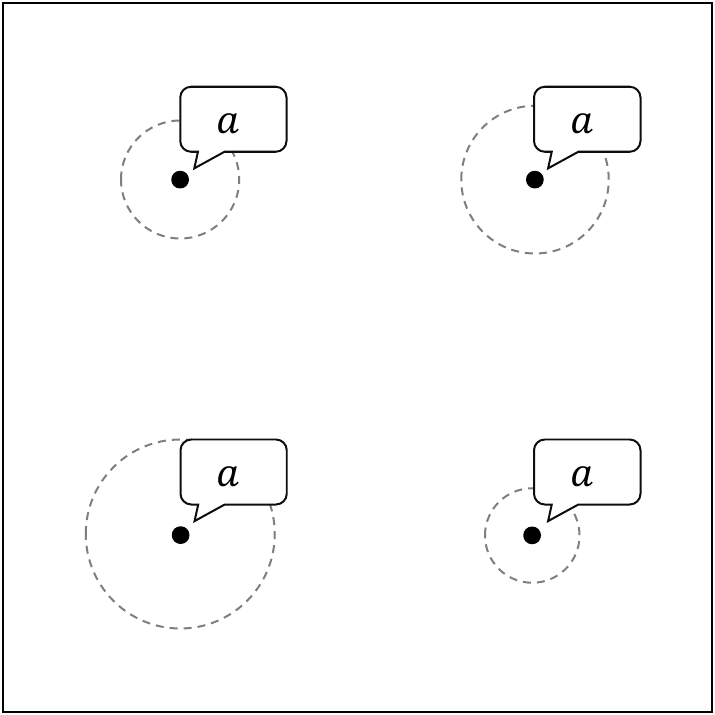}	\label{fig_example-5-a}}
        \subfigure[A distribution in which space-oriented partition is not effective]{%
			\includegraphics[width=0.485\linewidth]{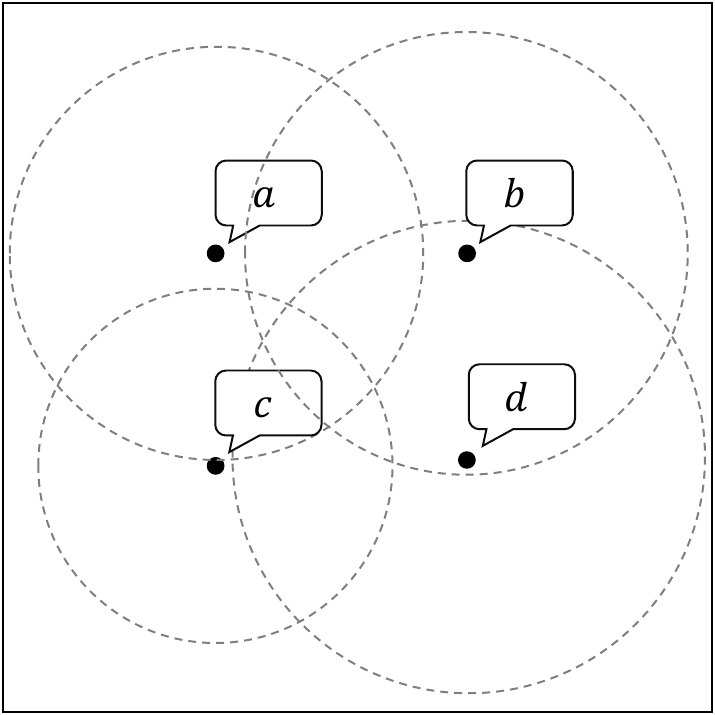}	\label{fig_example-5-b}}
        \caption{An example that depicts data distributions for considering a better partitioning approach. Black points, dashed circles, and balloons represent the locations of subscriptions $s$, $B(s)$, and keywords of $s$, respectively.}
        \label{figure_example-5}
	\end{center}
\end{figure}

\textbf{Motivation.}
Our baselines partition $S$ based only on either keyword or spatial information.
However, given a subspace, a better partition approach is dependent on the space and keyword distributions of the subspace.
For example:

\vs
\noindent
\textsc{Example 5.}
\textit{Figure \ref{figure_example-5} depicts two data distributions.
Black points, dashed circles, and balloons represent the locations of subscriptions $s$, $B(s)$, and keywords of $s$, respectively.}

\textit{Focus on Figure \ref{fig_example-5-a} and let the solid rectangle show $r$.
We see that $B(s)$ of each subscription $s$ is small, thereby the entire cost is small if we use space-oriented partition for $r$, because the pruning probability becomes large.
Next, consider Figure \ref{fig_example-5-b}.
Each subscription has a large $B(s)$ and it overlaps with the others.
For this distribution, space-oriented partition is clearly not a good choice, because the size of the rectangle that encloses each ball does not change much even if $r$ is partitioned.}

\vs
Motivated by the above observation, D$k$M-SKS considers a better partitioning approach when it partitions a (sub)set of subscriptions, to minimize the entire load.
Then D$k$M-SKS assigns each subscription to a specific worker based on the greedy algorithm \cite{graham1969bounds} and an additional heuristic.

\subsection{Cost Estimation}
D$k$M-SKS also utilizes $O_{init}$ to estimate the cost of a subscription.
Different from keyword- and space-oriented partition, D$k$M-SKS considers both space and keyword information.
Consider $S_{r}$ a subset of $S$, and we have a rectangle $R$ that encloses the balls of subscriptions in $S_{r}$.
Based on a similar idea to Equation (\ref{equation_space-prob}), the probability that an object $o$ is generated within $R$ and there exist the other objects in $R$ which contain a keyword $\lambda \in o.\psi$ is
\begin{equation}
	P(R,\lambda) = \frac{|O_{R,\lambda}|}{|O_{init}|},	\label{equation_dkm-prob}
\end{equation}
where $O_{R,\lambda}$ is a set of objects $\in O_{init}$ that exist within $R$ and contain $\lambda$ in their keywords.
Now take a subscription $s \in S_{r}$.
Equation (\ref{equation_dkm-prob}) focuses on a single keyword, thereby an estimated cost of $s$ is
\begin{equation}
	C(s) = \sum_{\lambda \in s.\psi}P(R,\lambda).	\label{equation_dkm-cost}
\end{equation}
Then the cost of $S_{r}$ is defined as
\begin{equation}
	C(S_{r}) = \sum_{s \in S_{r}}C(s).	\label{equation_dkm-cost_}
\end{equation}
Note that D$k$M-SKS provides a cost both with a single subscription and a set of subscriptions.

\subsection{Subscription Partition and Assignment}
\underline{\textbf{Subscription partition.}}
Given $S_{r}$, D$k$M-SKS selects a better approach to $S_{r}$ by considering the following two partitioning approaches.

\vs
\noindent
\textsc{Space-only-Partition.}
Consider a space $r$ where $S_{r}$ exists.
This approach partitions $r$ into equal-sized disjoint subspaces $r_{i}$ ($1 \leq i \leq 4$), as with space-oriented partition.
Then D$k$M-SKS obtains a set of subscriptions $S_{r_{i}}$ that exist in $r_{i}$.

\vs
\noindent
\textsc{Hybrid-Partition.}
This approach also partitions $S_{r}$ into four disjoint subsets $S_{h_{i}}$ ($1 \leq i \leq 4$), but in a different way from \textsc{Space-only-Partition}.
(The reason why this approach obtains four subsets is to be comparable to \textsc{Space-only-Partition}.)
This approach utilizes $C(s)$, which is defined in Equation (\ref{equation_dkm-cost}), and considers both space and keyword information.
More specifically, D$k$M-SKS sorts $S_{r}$ in descending order of $C(s)$, and runs the greedy algorithm to assign each $s \in S_{r}$ to $S_{h_{i}}$ with the minimum $\sum_{s \in S_{h_{i}}}C(s)$.

\vs
\noindent
We do not consider keyword-only partition, because it does not consider spatial information and cannot reduce the size of $R$.

We define a better partition as the one with less entire cost than the other after $S_{r}$ is partitioned.
The entire costs $C_{s}$ and $C_{h}$, which are respectively provided by \textsc{Space-only-Partition} and \textsc{Hybrid-Partition}, are defined as
\begin{equation*}
	C_{s} = \sum_{1 \leq i \leq 4}C(S_{r_{i}})
\end{equation*}
and
\begin{equation*}
	C_{h} = \sum_{1 \leq i \leq 4}C(S_{h_{i}}).
\end{equation*}
If $C_{s} < C_{h}$, D$k$M-SKS selects \textsc{Space-only-Partition}.
Otherwise, D$k$M-SKS selects \textsc{Hybrid-Partition}.

\vs
\noindent
\textbf{Algorithm description.}
Now we are ready to introduce how to partition $S$ through D$k$M-SKS.
Algorithm \ref{algo_partition} describes the detail.
The objective of this algorithm is to obtain at least $\gamma_{1} \cdot m$ subsets of $S$, where $\gamma_{1}$ is a system parameter.

For ease of presentation, assume that we are given a subset $S'$ of $S$.
(At initialization, $S' = S$.)
D$k$M-SKS considers partitioning $S'$.
Because \textsc{Hybrid-Partition} needs to compute Equation (\ref{equation_dkm-cost}) for each subscription in $S'$, it incurs a large computational cost if $|S'|$ is large.
Therefore, if $|S'| > \gamma_{2}$, where $\gamma_{2}$ is also a system parameter, D$k$M-SKS always utilizes \textsc{Space-only-Partition} to partition $S'$ into four subsets.
On the other hand, if $|S'| \leq \gamma_{2}$, D$k$M-SKS tests both \textsc{Space-only-Partition} and \textsc{Hybrid-Partition}.
D$k$M-SKS then selects the result of \textsc{Space-only-Partition} if $C_{s} < C_{h}$.
Otherwise, D$k$M-SKS selects that of \textsc{Hybrid-Partition}.
The four subsets obtained by a better partition are inserted into a collection $\mathcal{S}$ of subsets.
After that, $\mathcal{S}$ is sorted in descending order of the estimated cost of subset.
D$k$M-SKS checks $|\mathcal{S}|$, and if $|\mathcal{S}| < \gamma_{1} \cdot m$, D$k$M-SKS picks the subset with the largest cost and repeats the above operations.

\begin{algorithm}[!t]
	\caption{\textsc{Subscription-Partitioning}}
    \label{algo_partition}
	\DontPrintSemicolon
	\KwIn {$S$ (a set of subscriptions), $m$ workers, $\gamma_{1}$, and $\gamma_{2}$ (system parameters)}
	$\mathcal{S} \leftarrow \langle S,0\rangle$\;
    \While {$|\mathcal{S}| < \gamma_{1} \cdot m$}{
    	$\langle S',C(S')\rangle \leftarrow$ the front of $\mathcal{S}$\;
        $\mathcal{S} \leftarrow \mathcal{S} - \langle S',C(S')\rangle$\;
        \eIf {$|S'| > \gamma_{2}$}{
        	$\mathbb{S} \leftarrow$ \textsc{Space-only-Partition}$(S')$\;
            \For {each $S_{r} \in \mathbb{S}$}{
            	$\mathcal{S} \leftarrow \mathcal{S} \cup \langle S_{r},C(S_{r})\rangle$\;
            }
        }
        {
        	$\mathbb{S} \leftarrow$ \textsc{Space-only-Partition}$(S')$\;
            $\mathbb{S'} \leftarrow$ \textsc{Hybrid-Partition}$(S')$\;
            \eIf {$C_{s} < C_{h}$}{
            	\For {each $S_{r} \in \mathbb{S}$}{
            		$\mathcal{S} \leftarrow \mathcal{S} \cup \langle S_{r},C(S_{r})\rangle$\;
            	}
            }{
            	\For {each $S_{h} \in \mathbb{S'}$}{
            		$\mathcal{S} \leftarrow \mathcal{S} \cup \langle S_{h},C(S_{h})\rangle$\;
            	}
            }
        }
        Sort $\mathcal{S}$ in descending order of $C(S')$
    }
    \textbf{return} $\mathcal{S}$
\end{algorithm}
\begin{algorithm}[!t]
	\caption{\textsc{Subscription-Assignment} for D$k$M-SKS}
    \label{algo_assignment_}
	\DontPrintSemicolon
	\KwIn {$\mathcal{S}$ (a collection of subsets of $S$) and $m$ workers}
    Set $C(w) = 0$ for each worker\;
	Sort $\mathcal{S}$ in descending order of cost\;
    \For {each $S' \in \mathcal{S}$}{
    	Sort $S'$ in descending order of cost\;
        \For {each $s \in S'$}{
            $w \leftarrow \argmin_{m}C(w)$\;
        	$S(w) \leftarrow S(w) \cup \{s\}$, $C(w) \leftarrow C(w) + C(s)$
        }
    }
\end{algorithm}

\vs
\noindent
\underline{\textbf{Subscription assignment.}}
From the subscription partition, D$k$M-SKS has a collection $\mathcal{S}$ of subsets $S'$.
It is important to note that $S'$ tends to contain subscriptions with close locations and similar keyword sets.
That is, given a new object $o$, the $k$NN of all subscriptions in $S'$ may change by $o$.
In this case, assigning each subscription $s \in S'$ to a different worker is better than assigning $S'$ to a worker, to exploit the parallel $k$NN update.
We use this heuristic for subscription assignment.

\vs
\noindent
\textbf{Algorithm description.}
D$k$M-SKS employs a similar approach to keyword-oriented partition for subscription assignment.
In other words, D$k$M-SKS uses the greedy algorithm \cite{graham1969bounds}, but how to access subscriptions is different.
Given $\mathcal{S}$, we first sort $\mathcal{S}$ in descending order of cost obtained from Equation (\ref{equation_dkm-cost_}).
Then, for each $S' \in \mathcal{S}$, we sort $S'$ as with $\mathcal{S}$ and run the greedy algorithm.
Algorithm \ref{algo_assignment_} elaborates this operation.

\subsection{$k$NN Update Algorithm}
Actually, D$k$M-SKS can employ an arbitrary index for updating $k$NN of each subscription.
This is a good property because it can always make use of a state-of-the-art.
By default, in D$k$M-SKS, each worker $w$ utilizes a hybrid structure of a grid and an inverted file, because this structure is update-friendly.
The grid is a set of cells, and for each cell, we implement an inverted file.
More specifically, consider a subscription $s \in S(w)$ and a cell $g$ that overlaps with a ball of $s$, $B(s)$.
This cell $g$ maintains $s$ by its inverted file $g.I$ (i.e., $g.I[\lambda]$ maintains $s$ if $\lambda \in s.\psi$).

Given a new object $o$ broadcast by the main server, each worker $w$ obtains the cell $g$ to which $o$ is mapped.
Then $w$ considers whether or not it needs to update the $k$NN of subscriptions in $S(w)$ from $g.I$ (i.e., subscriptions that do not contain any keywords in $o.\psi$ are pruned).
If necessary, $w$ updates the $k$NN of corresponding subscriptions, then updates $g.I$ accordingly.

\begin{figure}[!t]
	\begin{center}
        \subfigure[Subscription partitioning of D$k$M-SKS]{%
			\includegraphics[width=0.80\linewidth]{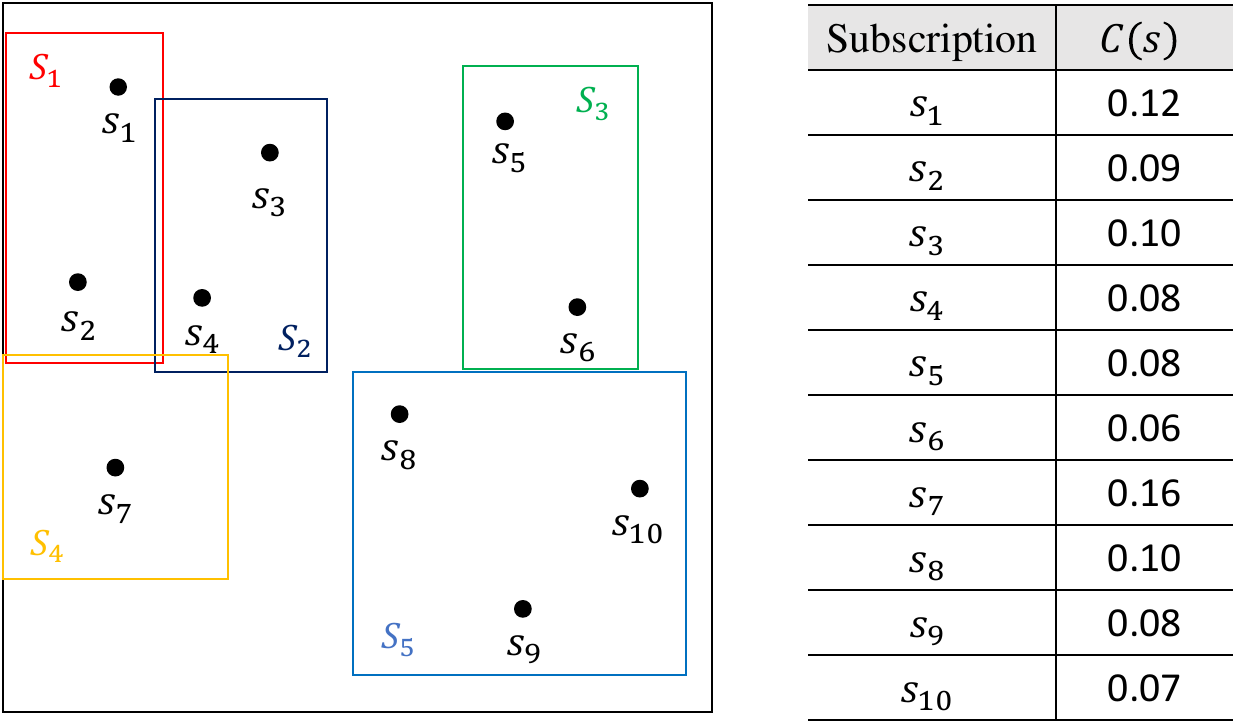}	\label{fig_example-6-a}}
        \subfigure[Subscriptions assigned to $w_{1}$ and the data structure maintained by $w_{1}$]{%
			\includegraphics[width=0.99\linewidth]{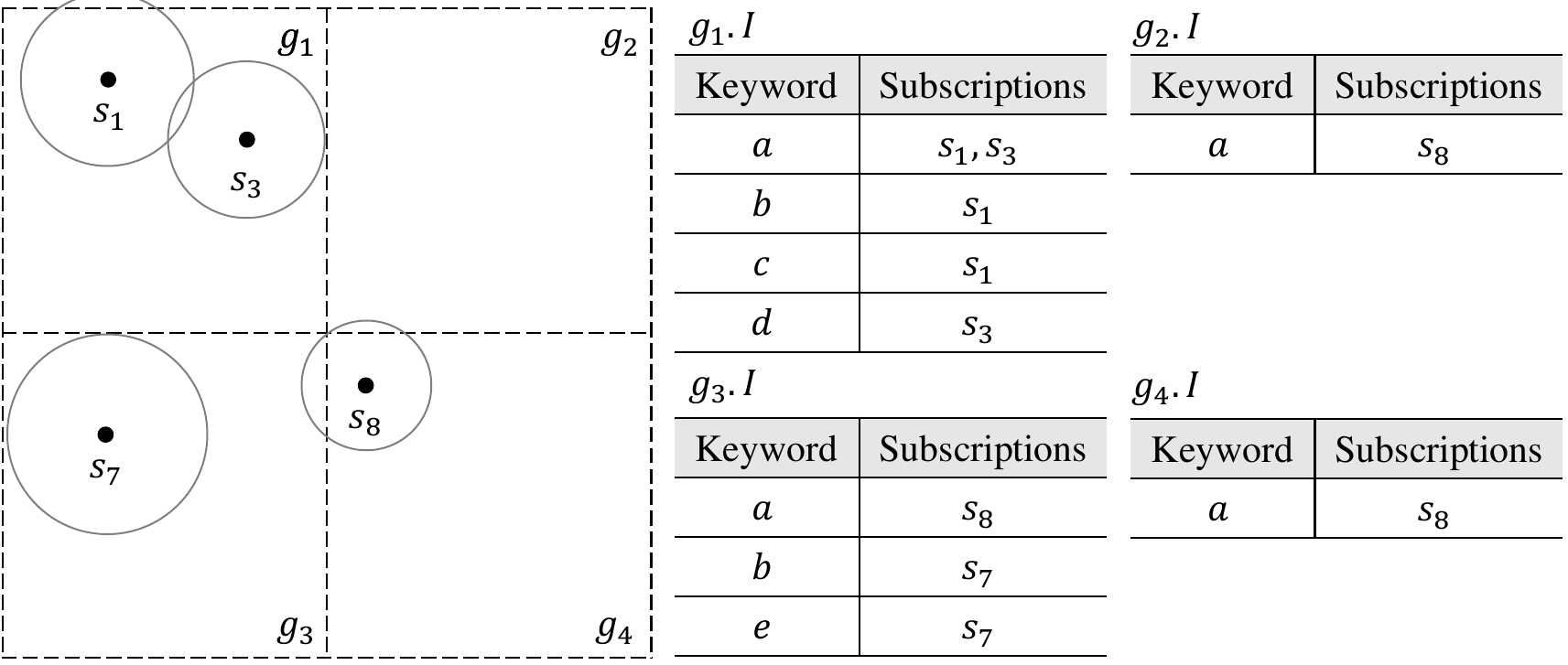}	\label{fig_example-6-b}}
        \subfigure[Subscriptions assigned to $w_{2}$ and the data structure maintained by $w_{2}$]{%
			\includegraphics[width=0.99\linewidth]{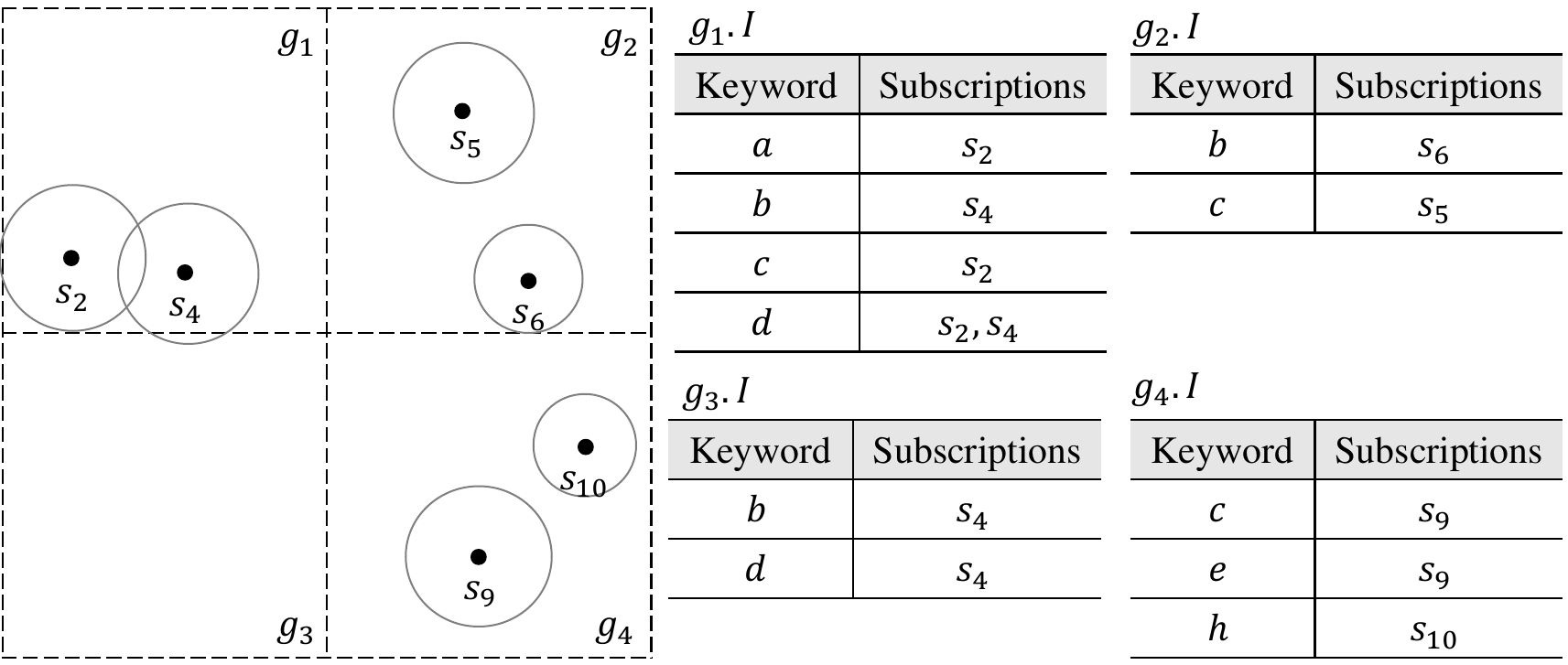}	\label{fig_example-6-c}}
        \caption{An example of subscription partitioning and assignment of D$k$M-SKS}
        \label{figure_example-6}
	\end{center}
\end{figure}

\vs
\noindent
\textsc{Example 6.}
\textit{D$k$M-SKS partitions $S$ in Figure \ref{figure_example-2} into two disjoint subsets for two workers $w_{1}$ and $w_{2}$.
Assume that the table in Figure \ref{fig_example-6-a} depicts the estimated cost of each subscription in D$k$M-SKS.
Assume furthermore that the result of partitioning is $\{S_{1},S_{2},S_{3},S_{4},S_{5}\}$, which is also shown in Figure \ref{fig_example-6-a}.
Following Algorithm \ref{algo_assignment_}, the result of subscription assignment of D$k$M-SKS is $\{s_{1},s_{3},s_{7},s_{8}\}$ for $w_{1}$ and $\{s_{2},s_{4},s_{5},s_{6},s_{9},s_{10}\}$ for $w_{2}$, which are respectively illustrated in Figures \ref{fig_example-6-b} and \ref{fig_example-6-c}.}

\textit{Consider a case where a new object $o_{10}$ (see Figure \ref{figure_example-4}), which contains keyword \textsf{b}, is generated.
Since it is mapped to $g_{3}$, $w_{1}$ needs to consider $s_{7}$, which can be seen from $g_{3}.I[\textsf{b}]$.
Similarly, $w_{2}$ needs to consider $s_{4}$.
Compared with Example 4, D$k$M-SKS shows a better load balancing.}

\subsection{Dealing with New Subscriptions}
Recall that the estimated cost of a subscription $s$ is obtained from $S_{r} \subset S$ ($s \in S_{r}$), as described in Equations (\ref{equation_dkm-prob}) and (\ref{equation_dkm-cost}).
When a new subscription $s_{n}$ is registered, it does not belong to any subset of $S$.
A straightforward approach to providing an estimated cost with $s_{n}$ is to re-conduct Algorithm \ref{algo_partition}.
It is obvious that this approach is very computationally expensive.
However, it is desirable that $s_{n}$ is handled as if $s_{n}$ has been registered at the initial time.
We therefore take an approximate approach to estimating the cost of $s_{n}$.

Consider a collection $\mathcal{S}$ of subsets of $S$ obtained by Algorithm \ref{algo_partition}.
The main server maintains $R$ for each subset $S_{r} \in \mathcal{S}$.
Given a new subscription $s_{n}$, we first do the same operation as space-oriented partition: the main server computes its $k$NN among $O_{init}$, and then computes the rectangle $R_{n}$ that encloses $B(s_{n})$.
Let $R \cap R_{n}$ be the overlapping area between $R$ and $R_{n}$.
Now $|R \cap R_{n}|$ can be the overlapped area size.
The main server computes
\begin{equation}
	S^{*} = \argmax_{S_{r} \in \mathcal{S}}|R \cap R_{n}|.	\label{equation_dkmsks-new}
\end{equation}
(In practice, $|\mathcal{S}|$ is small, so the cost of this computation is trivial.)
Let $R^{*}$ be the rectangle of $S^{*}$, and $R^{*}$ is the rectangle that overlaps with $R_{n}$ the most.
By using Equation (\ref{equation_dkm-cost}) with $R^{*}$, $s_{n}$ obtains its estimated cost.
Then $s_{n}$ is assigned to the worker $w$ with the minimum cost $C(w)$.

\section{Experiment}	\label{section_experiment}

\subsection{Setting}
We conducted experiments on a cluster of six machines.
One of them is a main server with 3.0GHz Intel Xeon Gold with 512GB RAM.
The others are equipped with 6-core 2.4GHz Intel Core i7-8700T and 32GB RAM.
We used one core as a worker.
The main server and workers communicate via a 1Gbps Ethernet network.

As with \cite{wang2016skype}, we set $|O_{init}| = 1,000,000$.
That is, when 1,000,000 objects were generated, we partitioned $S$ into $m$ workers.
After that, we generated 1,000 objects and requested 100 subscription insertions and deletions \textit{for each time-stamp}.

\vs
\noindent
\textbf{Dataset.}
We used two real spatial-keyword stream datasets, Place \cite{place} and Twitter \cite{twitter}.
Table \ref{table_statictics} shows the statistics of these datasets.
We generated subscriptions for each dataset, so that they follow the distributions of the corresponding dataset \cite{wang2017top}.
When a subscription $s$ was generated, we randomly picked one object to determine its location $s.p$ and then picked at most five keywords at uniformly random from its keyword set to obtain $s.\psi$.
The value of $s.k$ was a random integer $\in [1,k_{max}]$.

\vs
\noindent
\textbf{Algorithm.}
We evaluated the following algorithms:
\begin{itemize}
	\setlength{\leftskip}{-4.0mm}
	\item	\textit{PS$^2$Stream} \cite{chen2017distributed}: a state-of-the-art algorithm for continuous spatial-keyword \textit{range} queries.
    		We extend the original algorithm so that it can deal with our problem.
    \item	\textit{KOP}: our first baseline, keyword-oriented partition, introduced in Section \ref{section_keyword}.
    \item	\textit{SOP}: our second baseline, space-oriented partition, introduced in Section \ref{section_space}.
    \item	\textit{D$k$M-SKS}: our proposed solution in this paper.
\end{itemize}
All algorithms were implemented in C++.

\vs
\noindent
\textbf{Parameter.}
The default values of $m$, $k_{max}$, and the initial $|S|$ are 20, 10, and 10,000,000, respectively.
When we investigated the impact of a given parameter, the other parameters were fixed.
In addition, we set $\theta = 20$, $\gamma_{1} = 100,000$, and $\gamma_{2} = 20$ from preliminary experiments.

\begin{figure*}[!t]
	\begin{center}
    \includegraphics[width=0.375\linewidth]{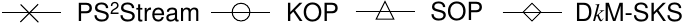}
    
        \subfigure[Place]{%
		\includegraphics[width=0.49\linewidth]{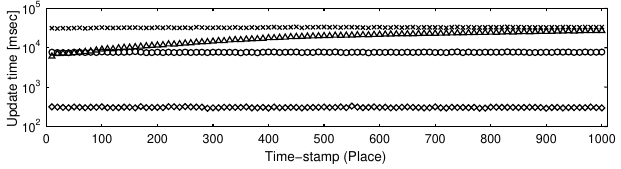}	\label{fig_time_place}}
        \subfigure[Twitter]{%
		\includegraphics[width=0.49\linewidth]{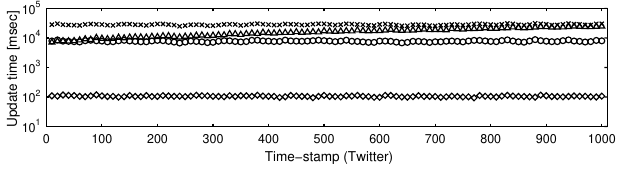}	\label{fig_time_twitter}}
        \vspace{-5.0mm}
        \caption{Update time as a function of time}
        \label{figure_time}
	\end{center}
\end{figure*}

\vs
\noindent
\textbf{Criteria.}
To evaluate the performance of each algorithm, we measured the following criteria:
\begin{itemize}
	\setlength{\leftskip}{-4.0mm}
	\item	Update time: this is the average computation time for updating $k$NN objects of all registered subscriptions and time for dealing with subscription insertions and deletions, per time-stamp.
    \item	Load balance: this is the average difference between the maximum and the minimum time to finish $k$NN update between workers, per time-stamp.
\end{itemize}

\begin{table}[!t]
\begin{center}
	\caption{Dataset statistics}
    \label{table_statictics}
    \vspace{-3.0mm}
	\begin{tabular}{c||c|c} \hline
                Dataset								& Place		& Twitter		\\ \hline \hline
        		Cardinality							& 9,356,750	& 20,000,000	\\ \hline
                Distinct \# of keywords				& 53,931	& 2,225,654		\\ \hline
                Avg. \# of keywords in one object	& 2.94		& 5.51			\\ \hline
	\end{tabular}
\end{center}
\end{table}

\subsection{Results}
\textbf{Justification of using $O_{init}$ and analysis.}
We first empirically demonstrate that cost estimation based on $O_{init}$ functions well.
Figure \ref{figure_time} depicts the time-series of the update time of each algorithm.
The result of D$k$M-SKS shows that its update time does not vary even as new objects are given, new subscriptions are inserted, and some subscriptions are removed, on both Place and Twitter.
This suggests that D$k$M-SKS keeps load balancing, and its cost estimation yields this result.
Table \ref{table_load} depicts that the load of each worker in D$k$M-SKS is actually balanced.

We see that PS$^2$Stream also has this tendency.
However, this result does not mean that PS$^2$Stream has load balancing.
We observed that the initial partition of PS$^2$Stream incurs very imbalance partition, i.e., one worker $w$ has a very heavy load at initialization.
Because of this, new subscriptions are assigned to the other workers, but this does not overcome the load imbalance.
Therefore, the update time of PS$^2$Stream is simply affected by the load of $w$.
This is also confirmed from Table \ref{table_load}, which shows that the load in PS$^2$Stream is significantly imbalance.

Next, we focus on KOP.
This algorithm also has a similar result.
From Figure \ref{figure_time} and Table \ref{table_load}, we see that the load balance of KOP is not so bad.
Because both subscription partition and $k$NN update in KOP consider only keyword information, they go well together.
However, KOP is outperformed by D$k$M-SKS, which considers both spatial and keyword information.
This result confirms the effectiveness of the partitioning approach in D$k$M-SKS.

Let us consider SOP here.
Figure \ref{figure_time} shows that, different from the other algorithms, the update time of SOP increases, as time goes by.
This is derived from low accuracy of its cost estimation for new subscriptions.
Specifically, given a new subscription, its cost estimated by SOP is usually small, although it is large in practice.
Because of this, the load of a worker, which has new subscriptions, becomes heavy and bottleneck of the system.
Table \ref{table_load} also demonstrates this fact.

\begin{table}[!t]
\begin{center}
	\caption{Load balance [msec] (default parameters)}
    \label{table_load}
    \vspace{-3.0mm}
	\begin{tabular}{c||c|c|c|c} \hline
                Algorithm	& PS$^2$Stream	& KOP		& SOP		& D$k$M-SKS	\\ \hline \hline
        		Place		& 27490.11		& 549.23	& 14037.33	& 32.08		\\ \hline
                Twitter		& 21962.50		& 486.97	& 12265.40	& 51.03		\\ \hline
	\end{tabular}
\end{center}
\end{table}
\begin{table}[!t]
\begin{center}
	\caption{Decomposed time [msec] on Place}
    \label{table_decomposed-time}
    \vspace{-3.0mm}
	\begin{tabular}{c||c|c|c|c} \hline
                Algorithm			& PS$^2$Stream	& KOP		& SOP		& D$k$M-SKS	\\ \hline \hline
        		$k$NN update		& 32892.90		& 7684.48	& 18549.04	& 471.00	\\ \hline
                Subscription ins.	& 1.51			& 1.59		& 1110.65	& 35.36		\\ \hline
                Subscription del.	& 1.54			& 1.42		& 0.79		& 1.10		\\ \hline
	\end{tabular}
\end{center}
\end{table}

Last, we investigate the detail of update time.
Table \ref{table_decomposed-time} decomposes the update time of each algorithm on Twitter into $k$NN update time, subscription insertion time, and subscription deletion time, each of which includes index update time.
The result on Twitter is omitted, because its tendency is similar to that on Place.
It can be seen that the main part of the update time is $k$NN update time, and subscription deletion needs a trivial cost.
D$k$M-SKS significantly outperforms (is more than 10 times faster than) the other algorithms and exploits available workers to reduce $k$NN update time.
We see that the query insertion time of D$k$M-SKS is slower than those of KOP and PS$^2$Stream.
This is because D$k$M-SKS needs to compute Equation (\ref{equation_dkmsks-new}), which incurs a more cost than Equation (\ref{equation_space-prob}).
Also, we can observe that the subscription insertion time of SOP is much longer than those of the others.
As explained earlier, (most) new subscriptions are assigned to a single worker $w$.
Hence $w$ incurs a long index update time.

\begin{figure*}[!t]
	\begin{center}
    \includegraphics[width=0.25\linewidth]{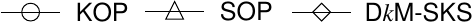}
    
        \subfigure[Update time (Place)]{%
		\includegraphics[width=0.24\linewidth]{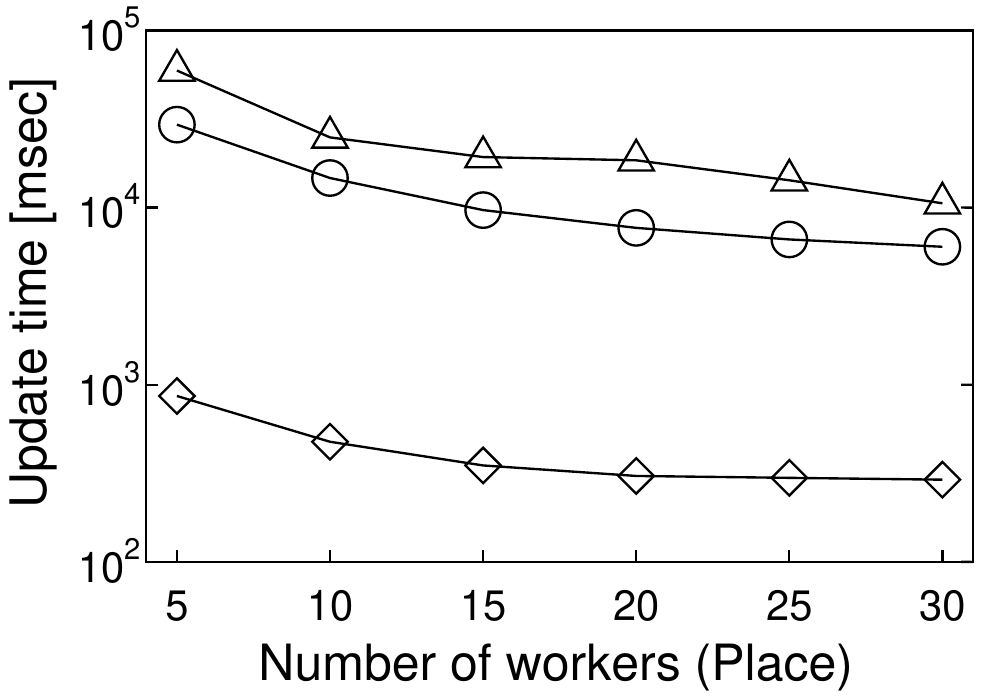}	\label{fig_m_time_place}}
        \subfigure[Update time (Twitter)]{%
		\includegraphics[width=0.24\linewidth]{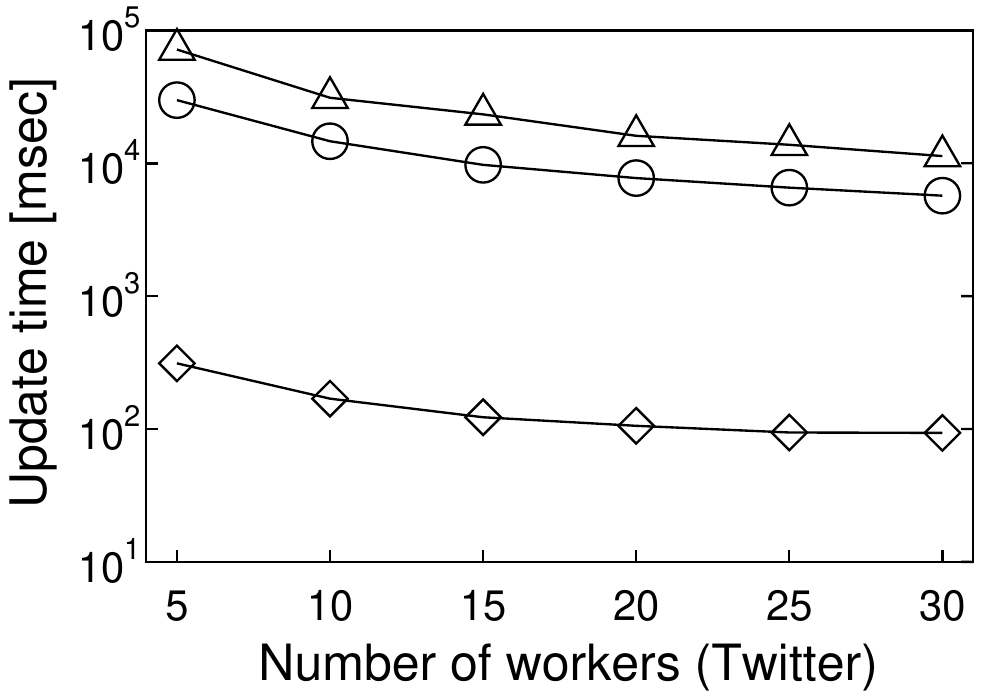}	\label{fig_m_time_twitter}}
        \subfigure[Load balance (Place)]{%
		\includegraphics[width=0.24\linewidth]{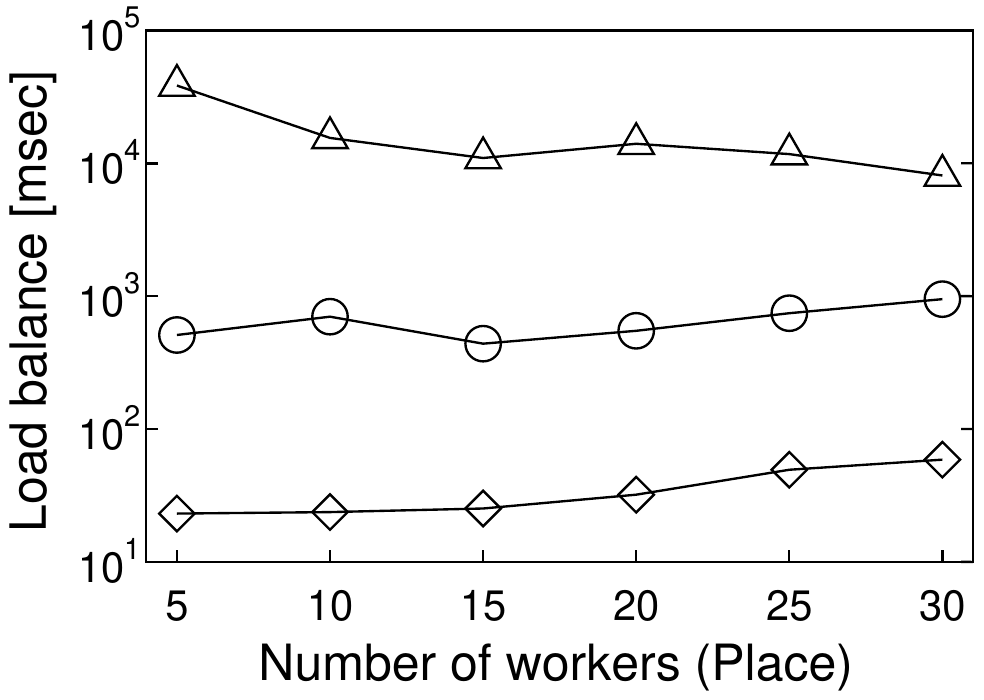}	\label{fig_m_load_place}}
        \subfigure[Load balance (Twitter)]{%
		\includegraphics[width=0.24\linewidth]{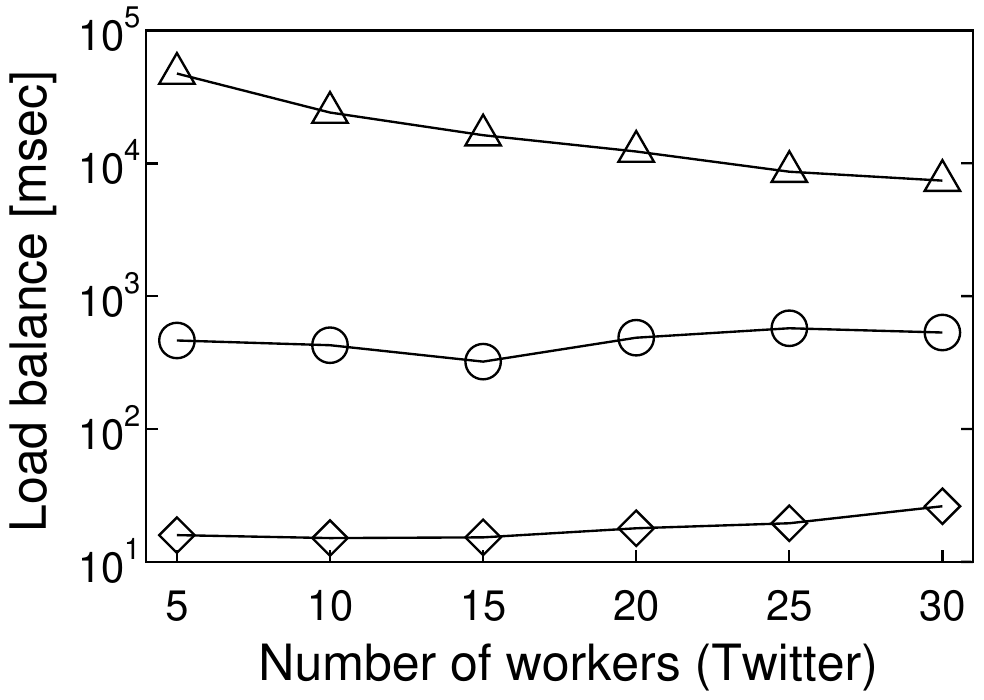}	\label{fig_m_load_twitter}}
        \vspace{-3.0mm}
        \caption{Impact of $m$}
        \label{figure_m}
        \vspace{-4.0mm}
	\end{center}
\end{figure*}

\vs
\noindent
\textbf{Varying $m$.}
We next study the impact of $m$, the number of workers.
Figure \ref{figure_m} depicts the result.
Since PS$^2$Stream is significantly outperformed by D$k$M-SKS, we omit its result.

We see that each algorithm reduces its update time as $m$ increases.
This is an intuitive result, since subscriptions are distributed to more workers.
The load balances of KOP and D$k$M-SKS are not affected by $m$ so much, since their cost estimations yield balanced load.
On the other hand, as $m$ increases, the load balance of SOP decreases.
The reason is simple: the update time of the worker with the largest load becomes shorter as $m$ increases.

\begin{figure*}[!t]
	\begin{center}
        \subfigure[Update time (Place)]{%
		\includegraphics[width=0.24\linewidth]{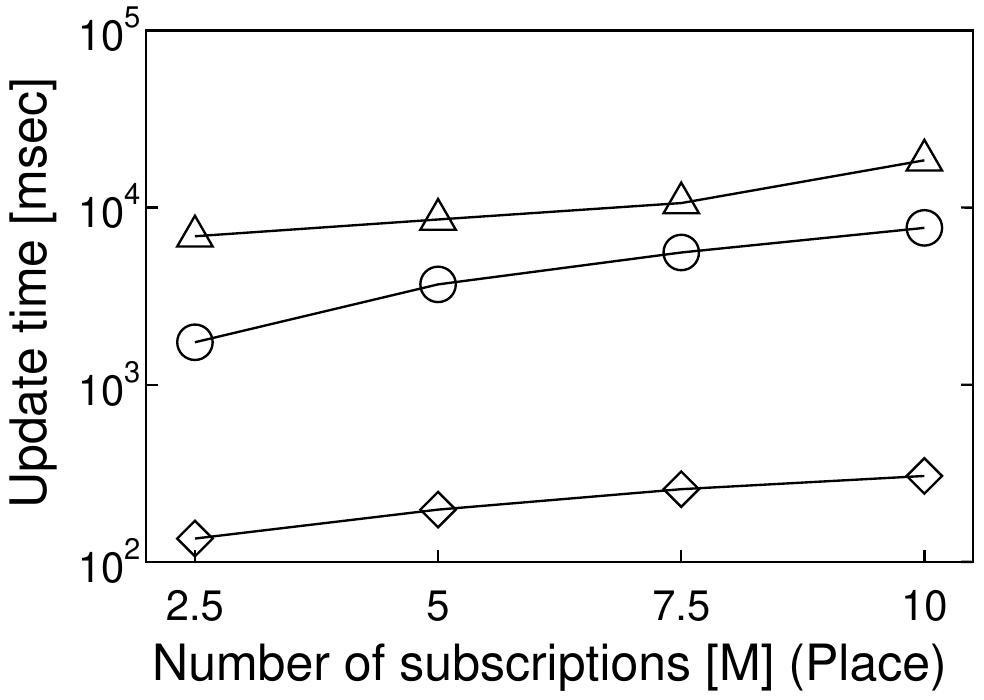}	    \label{fig_s_time_place}}
        \subfigure[Update time (Twitter)]{%
		\includegraphics[width=0.24\linewidth]{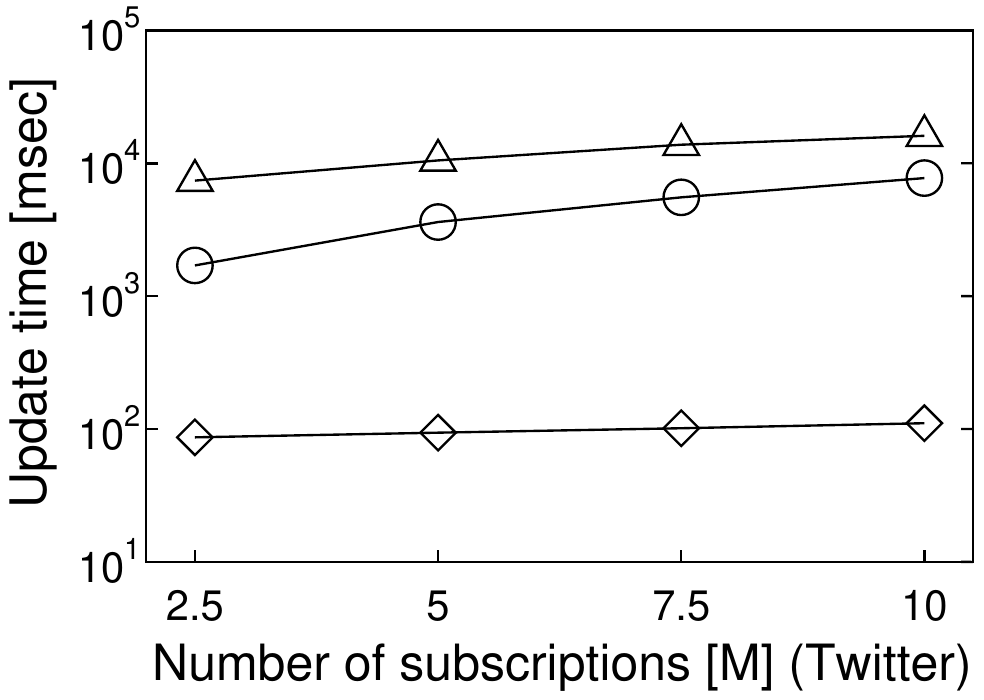}	\label{fig_s_time_twitter}}
        \subfigure[Load balance (Place)]{%
		\includegraphics[width=0.24\linewidth]{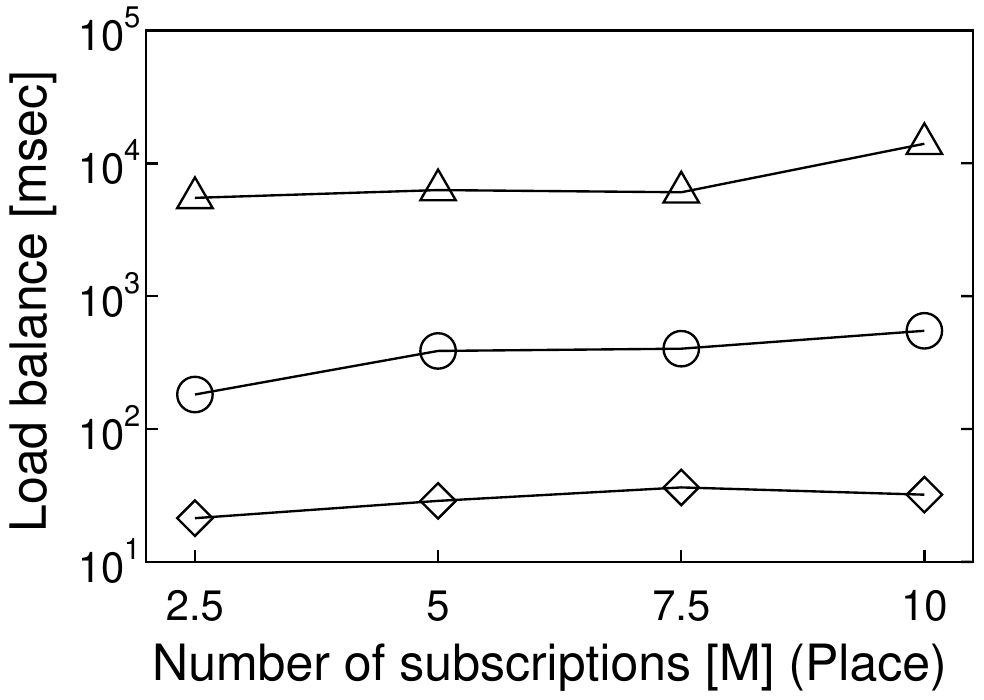}	    \label{fig_s_load_place}}
        \subfigure[Load balance (Twitter)]{%
		\includegraphics[width=0.24\linewidth]{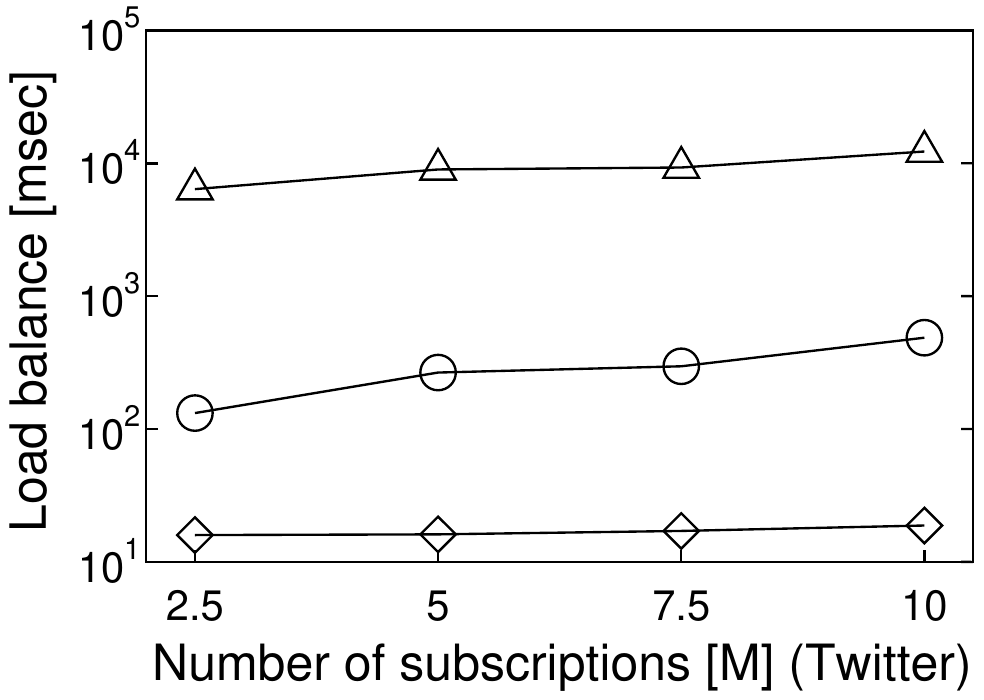}	\label{fig_s_load_twitter}}
        \vspace{-3.0mm}
        \caption{Impact of $|S|$}
        \label{figure_s}
        \vspace{-4.0mm}
	\end{center}
\end{figure*}
\begin{figure*}[!t]
	\begin{center}
        \subfigure[Update time (Place)]{%
		\includegraphics[width=0.24\linewidth]{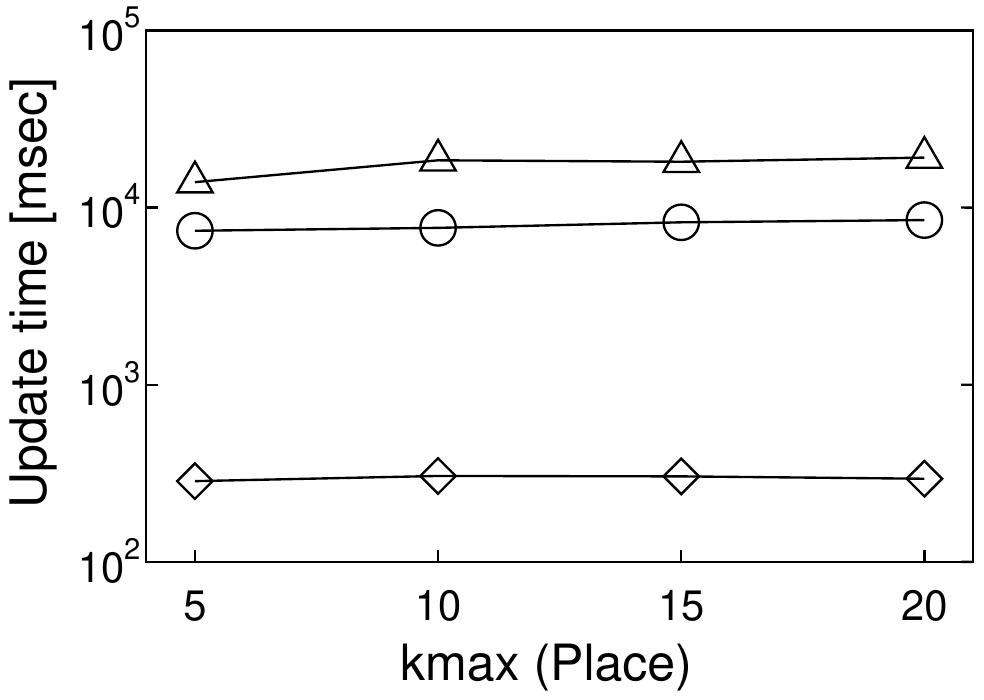}	    \label{fig_k_time_place}}
        \subfigure[Update time (Twitter)]{%
		\includegraphics[width=0.24\linewidth]{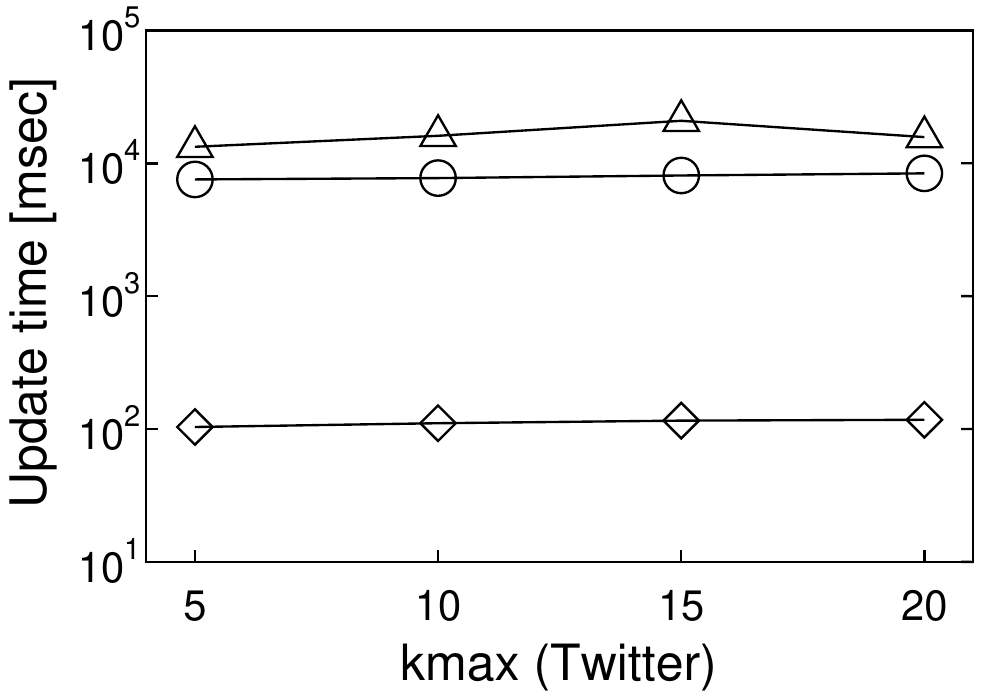}	\label{fig_k_time_twitter}}
        \subfigure[Load balance (Place)]{%
		\includegraphics[width=0.24\linewidth]{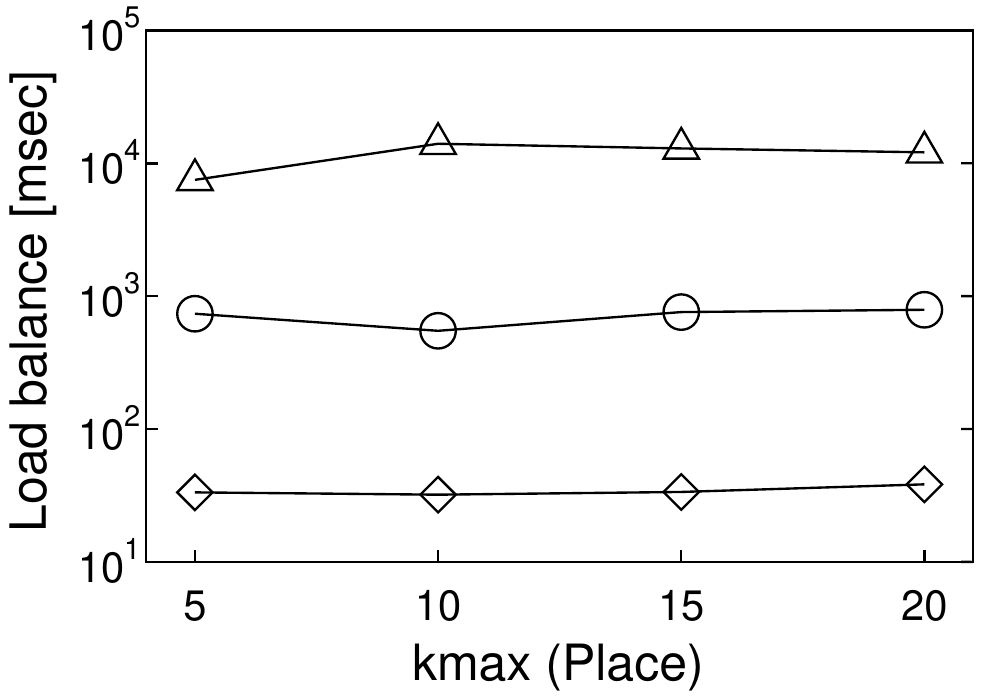}	    \label{fig_k_load_place}}
        \subfigure[Load balance (Twitter)]{%
		\includegraphics[width=0.24\linewidth]{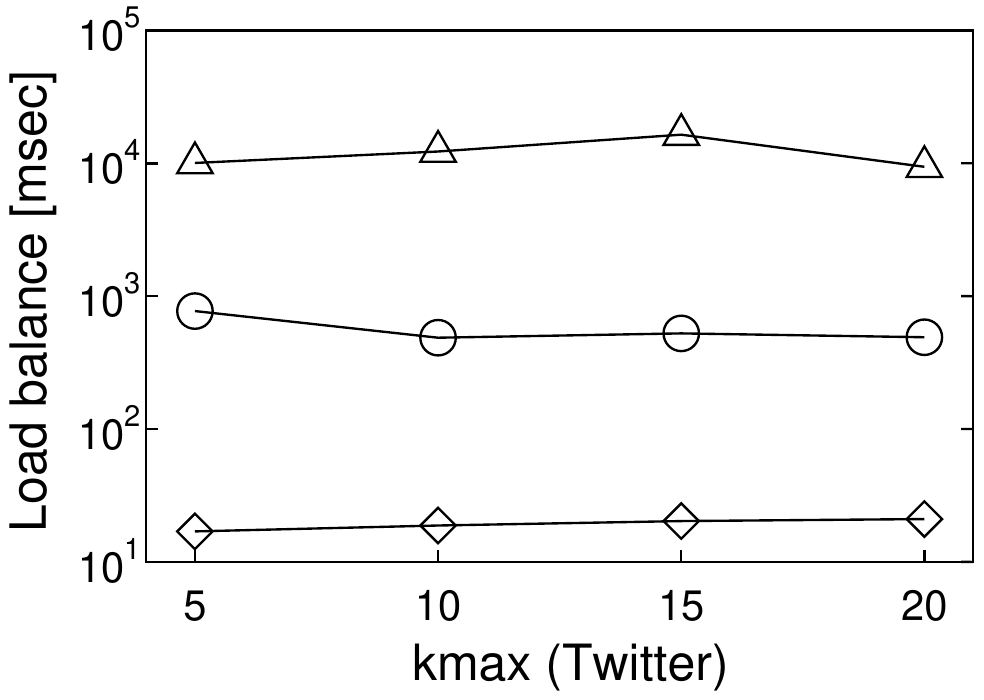}	\label{fig_k_load_twitter}}
        \vspace{-3.0mm}
        \caption{Impact of $k_{max}$}
        \label{figure_k}
        \vspace{-2.0mm}
	\end{center}
\end{figure*}

\vs
\noindent
\textbf{Varying $|S|$.}
To investigate the scalability of each algorithm, we studied the influence of $|S|$.
Figure \ref{figure_s} shows the result.
Due to the load imbalance, the update time of SOP is slow, even when $|S|$ is small.
KOP and D$k$M-SKS scale linearly to $|S|$, so D$k$M-SKS always outperforms KOP.
Note that the linear scalability shows that their subscription assignment functions well.
From Figs. \ref{fig_k_load_place} and \ref{fig_k_load_twitter}, we see that the load balance of each algorithm normally becomes larger, as $|S|$ increases.
It is however trivial increase.
For example, in D$k$M-SKS on Twitter, the difference is only 40[msec] between the cases of $|S| = 2.5 \cdot 10^{6}$ and $|S| = 10 \cdot 10^{6}$.

\vs
\noindent
\textbf{Varying $k_{max}$.}
Last, we study the impact of $k_{max}$, and the result is shown in Figure \ref{figure_k}.
The update time of KOP and D$k$M-SKS increases slightly as $k_{max}$ increases.
This is also a reasonable result, because, as $k_{max}$ increases, the probability that new objects update the $k$NN of some subscriptions becomes higher.
SOP shows a different pattern to KOP and D$k$M-SKS, because of its load imbalance.
It can be seen that the update time of SOP is simply affected by its load balance.

\section{Related Work}	\label{section_related-work}
\textbf{Spatial-keyword search.}
Due to the prevalence of spatial-keyword objects, algorithms for efficient searching them have been extensively devised \cite{chen2013spatial}.
For indexing a given dataset, these algorithms employ hybrid structures of spatial indices, e.g., R-tree and quadtree, and textual indices, such as inverted files \cite{cong2009efficient, zhang2016inverted}.
A famous index is IR-tree \cite{cong2009efficient}.
This is an essentially R-tree, but each node contains an inverted file.
The hybrid structure of grid and inverted file, which is employed by D$k$M-SKS, is derived from IR-tree.
An R-tree is not update-efficient, because it partitions data space based on a given dataset.
We therefore did not employ R-tree-like structures.
It is also important to notice that these works consider snapshot queries and static datasets, whereas we consider continuous queries and streaming data.

Some related queries support moving users \cite{wu2013moving}, find potential users \cite{choudhury2016maximizing}, and analyze the result of spatial-keyword search \cite{chen2015answering}.
These works are also totally different from our problem.

\vs
\noindent
\textbf{Distributed query processing system.}
Recently, distributed spatial query processing systems have been developed on Hadoop, Spark, and Storm.
(D$k$M-SKS is orthogonal to these systems.)
For example, Hadoop-GIS \cite{aji2013hadoop} and SpatialHaddop \cite{eldawy2015spatialhadoop} support efficient processing of spatial queries, e.g., range and $k$NN queries on MapReduce environments.
However, they do not consider keyword information and cannot deal with our problem.

Tornado \cite{mahmood2015tornado} is a system based on Storm \cite{storm} and supports spatio-textual queries.
The main focus of this system is to achieve efficient spatial-keyword query processing and not to support massive subscriptions.
Hence, it is not trivial for Tornado to provide subscription partitioning for continuous spatial-keyword $k$NN queries.
SSTD \cite{chen2020sstd} is also a system that supports spatio-textual queries on streaming data.
However, SSTD imposes, for objects, the condition that they have to contain all keywords specified by queries to match the queries.
This is too strict, resulting in no matching objects.

\vs
\noindent
\textbf{Location-aware Pub/Sub.}
There are many studies that addressed the problem of dealing with spatio-textual subscriptions.
Literatures \cite{li2013location, mahmood2018fast, mahmood2018adaptive, wang2015ap} considered continuous boolean \textit{range} queries as subscriptions.
Although PS$^2$Stream \cite{chen2017distributed} also deals with boolean range queries, this is the most related work to ours, because it also assumes the same distributed setting as ours.
Our empirical study has demonstrated that the cost model proposed in \cite{chen2017distributed} is not efficient for our problem and D$k$M-SKS significantly outperforms PS$^2$Stream.

Some studies \cite{chen2015temporal, nishio2017geo, nishio2020lamps, wang2016skype} also tackled the problem of spatio-textual $k$NN (or top-k) monitoring.
\cite{chen2015temporal} considers a decay model for streaming data, while \cite{nishio2020lamps, wang2016skype} do a sliding-window model.
In addition, they consider an aggregation function for object scoring, i.e., spatial proximity and keyword (textual) similarity are aggregated to a score through a weighting parameter $\alpha$.
Based on this scoring function, they monitor top-k objects for each subscription.
Their techniques are specific to this scoring function and their assumed streaming model (decay or sliding-window), thereby cannot deal with our problem.
Besides, it is well-known that specifying an appropriate $\alpha$ is generally hard for ordinary users \cite{he2012answering}.
We therefore consider boolean-based $k$NN monitoring, which is more user-friendly.

\section{Conclusion}	\label{section_conclusion}
In this paper, to scale well to massive objects and subscriptions in location-aware Pub/Sub environments, we proposed D$k$M-SKS, a distributed solution to the problem of spatial-keyword $k$NN monitoring of massive subscriptions.
D$k$M-SKS employs a new cost model to effectively reflect the load of a given subscription.
Besides, D$k$M-SKS partitions a set of subscriptions so that the entire load becomes as small as possible, then assigns each subscription to a specific worker while considering load balancing.
We conducted experiments on two real datasets, and the results demonstrate that D$k$M-SKS outperforms baselines and a state-of-the-art and scales well to massive subscriptions.

\section*{Acknowledgments}
This research is partially supported by JSPS Grant-in-Aid for Scientific Research (A) Grant Number 18H04095.

\bibliographystyle{abbrv}
\bibliography{sigproc}

\end{document}